\documentclass[aps,prc,fleqn,floatfix,twocolumn,twoside]{revtex4}
\usepackage{graphicx}
\usepackage{epstopdf}

\begin{document}

\title{Longitudinal decorrelation of anisotropic flows in heavy-ion collisions at the CERN Large Hadron Collider}

\author{Long-Gang Pang,$^1$ Guang-You Qin,$^1$ Victor Roy,$^{1,2}$ Xin-Nian Wang$^{1,3}$}
\affiliation{$^1$Institute of Particle Physics and Key Laboratory of Quarks and Lepton Physics (MOE),
Central China Normal University, Wuhan 430079, China}
\affiliation{$^2$Institute for Theoretical Physics, Goethe University, Max-von-Laue-Str.1, D-60438, Frankfurt am Main, Germany}
\affiliation{$^3$Nuclear Science Division, MS 70R0319, Lawrence Berkeley National Laboratory, Berkeley, CA 94720}

\author{Guo-Liang Ma}
\affiliation{Shanghai Institute of Applied Physics,Chinese Academy of Sciences, Shanghai 201800, China}

\date{\today}
\begin{abstract}

Fluctuations in the initial transverse energy-density distribution lead to anisotropic flows as observed in central high-energy heavy-ion collisions. Studies of longitudinal fluctuations of the anisotropic flows can shed further light on the initial conditions and dynamical evolution of the hot quark-gluon matter in these collisions. Correlations between anisotropic flows with varying pseudorapidity gaps in Pb+Pb collisions at the CERN Large Hadron Collider  are investigated using both an event-by-event (3+1)-D ideal hydrodynamical model with fluctuating initial conditions and the a multiphase transport (AMPT) Monte Carlo model for high-energy heavy-ion collisions.  Anisotropic flows at different pseudorapidities are found to become significantly decorrelated with increasing pseudo-rapidity gaps due to longitudinal fluctuations in the initial states of heavy-ion collisions. The longitudinal correlation of the elliptic flow shows a strong centrality dependence while the correlation of the triangular flow is independent of the centrality.  Longitudinal fluctuations as a source of the decorrelation are further shown to consist of a twist or gradual rotation in flow angles between the forward and backward direction and additional fluctuations on top of the twist.  Within the AMPT model, longitudinal correlations of anisotropic flows are also found to depend on the value of partonic cross sections. The implicatiosn of constraining the initial conditions and shear viscosity to entropy density ratio of the partonic matter in high-energy heavy-ion collisions are also discussed.

\end{abstract}
\pacs{25.75.Ld}
\maketitle
\section{INTRODUCTION}

One of the most important pieces of evidence for the formation of the quark-gluon plasma (QGP) in high-energy heavy-ion collisions at the BNL Relativistic Heavy-Ion Collider (RHIC) and the CERN Large Hadron Collider (LHC) is the strong collectivity of the bulk matter as measured through the anisotropic azimuth distributions of the final-state hadrons \cite{Adams:2003zg, Aamodt:2010pa, ATLAS:2011ah, Chatrchyan:2012ta}. The anisotropies are characterized in terms of the Fourier coefficients of the azimuth distributions which are  referred to as anisotropic flows or harmonic flows. Relativistic hydrodynamical models have been very successful in describing the observed anisotropic flows \cite{Heinz:2013th, Gale:2013da, Song:2013gia,Romatschke:2009im,Huovinen:2013wma} and in understanding the space-time evolution of the QGP fireball. Detailed comparisons between hydrodynamical model calculations and experimental data on anisotropic flows have provided unprecedented constraints on the transport properties, such as the shear viscosity to entropy density ratio, of the hot and dense QGP formed in high-energy heavy-ion collisions.

Early studies of anisotropic flows have mainly been focused on the second harmonic flow (elliptic flow) $v_2$ which originates from the almond shape of the produced fireball in the collision zone. Much attention has now been paid to fluctuations of the anisotropic flows of final hadrons due to the fluctuations in the initial states of heavy-ion collisions, such as the fluctuations of nucleon positions and color charges inside colliding nuclei and multiplicities of initial parton production \cite{Luzum:2013yya}. One of the most interesting consequences of initial-state fluctuations is the presence of elliptic flow in the most central nucleus-nucleus collisions, and finite odd harmonic flows in the final-state azimuthal distributions \cite{Alver:2010gr, Alver:2010dn, Petersen:2010cw, Staig:2010pn, Qin:2010pf, Ma:2010dv, Xu:2010du, Teaney:2010vd, Qiu:2011iv, Bhalerao:2011yg, Floerchinger:2011qf, Pang:2012uw, Gale:2012rq, Roy:2012pn, Pang:2013pma, Rybczynski:2013yba}. The elliptic, triangular and other higher-order harmonic flows have been measured in Au+Au collisions at RHIC and in Pb+Pb collisions at the LHC \cite{Adare:2011tg, Adamczyk:2013waa, ATLAS:2012at}. The anisotropic flows may also exist in small collision systems such as p+Pb collisions at the LHC \cite{Abelev:2012ola, Aad:2012gla, Chatrchyan:2013nka, Bozek:2013ska, Bzdak:2013zma, Qin:2013bha, Schenke:2014zha, Bzdak:2014dia}.

In the study of fluctuations of the initial-state geometry in terms of eccentricity vectors $\mathbf{\epsilon}\hspace{-5pt}\mathbf{\epsilon}_n = \epsilon_n \exp(i \Psi_n)$, both the magnitudes $\epsilon_n$ and the corresponding orientation angles (participant planes $\Psi_n$) fluctuate from event to event.
As a consequence of the hydrodynamic evolution of the fireball, anisotropic flows $v_n$ and the corresponding orientation angles (flow planes $\Phi_n$) fluctuate as well, where the flow vector is defined as $\mathbf{v}_n=v_n \exp{(i\Phi_n})$. The ATLAS and ALICE Collaborations have measured the event-by-event distributions of anisotropic flows $v_n$ for Pb+Pb collisions at the LHC \cite{Aad:2013xma, Timmins:2013hq}.
There have been studies on the correlations between event plane angles of different harmonics \cite{Qin:2011uw,Bhalerao:2011bp,Jia:2012sa,Qiu:2012uy, Bhalerao:2013ina,Teaney:2013dta}. Event-by-event fluctuations of anisotropic flows and their correlations have provided very important tools for studying the initial-state fluctuations and the transport properties of the produced QGP.

When measuring anisotropic flows and their correlations, one common practice is to correlate particles with a large pseudo-rapidity gap.
The use of a large pseudo-rapidity gap is to remove or minimize contributions from non-flow effects such as resonance decays and jet production.
Such a method is good under the assumption that the initial density distributions and final-state anisotropic flows at different rapidities are perfectly aligned (correlated), as in the case of boost invariance. However,  the density of the initial produced fireballs in real high-energy heavy-ion collisions  contains fluctuations not only in the transverse plane, but also in the longitudinal direction.  Such fluctuations are in addition to the fluctuations due to finite multiplicity in any given event which can be corrected using the sub-event method \cite{Poskanzer:1998yz,Ollitrault:1993ba}. Hydrodynamic calculations have shown that the inclusion of longitudinal fluctuations reduces the values of anisotropic flows \cite{Pang:2012he}. The longitudinal fluctuations may also lead to fluctuations of the final flow orientation angles at different pseudo-rapidities \cite{Petersen:2011fp, Xiao:2012uw, Jia:2014ysa, Jia:2014vja}. Therefore, detailed knowledge of longitudinal fluctuations and their manifestation in the final-state flow and correlation observables is essential in understanding the initial conditions of heavy-ion collisions and constraining transport properties of the QGP through comparisons between viscous hydrodynamic model calculations and experimental measurements.

In this work, we investigate longitudinal fluctuations and correlations of anisotropic flows in Pb+Pb collisions at the LHC. We use two different dynamical models to simulate the space-time evolution of the produced fireball: a multi-phase transport (AMPT) model and an event-by-event (3+1)-dimensional [(3+1)D] ideal relativistic hydrodynamical model. The fluctuating initial conditions in both models from HIJING simulations are taken to be identical. We study the correlations of anisotropic flows and their corresponding event-plane angles of the same order at different pseudo-rapidities. To investigate the sensitivity of the longitudinal correlations of anisotropic flows to the evolution dynamics, results from hydrodynamic model simulations are compared in detail with those from the AMPT model with different values of parton cross sections and durations of hadronic evolution. The remainder of this article is organized as follows. In the next section, we briefly discuss the (3+1)D ideal hydrodynamical model and the AMPT model that we use for simulations. Our main results for longitudinal correlations are presented in Sec. III.  A summary and some discussions are given in Sec. IV.

\section{Model Descriptions}
\label{sec:model}

\subsection{A (3+1)-dimensional ideal hydrodynamical model}

In the (3+1)-D ideal hydrodynamical model \cite{Pang:2012he} that we employ, initial conditions are obtained from the HIJING model \cite{Wang:1991hta,Gyulassy:1994ew,Wang:2009qb} (see the next subsection for the description of the AMPT model that uses the HIJING model as initial conditions for parton production), in which the fluctuations of initial energy and momentum densities in both the transverse plane and the longitudinal direction are taken into account. The initial energy-momentum tensor $T^{\mu\nu}$ is calculated from the momentum distribution of produced partons on a fixed proper time~($\tau_{0}$) surface using a Gaussian smearing (see Ref.~\cite{Pang:2012he} for details):

\begin{eqnarray}
{ T }^{ \mu \nu  }\left( { \tau  }_{ 0 },x,y,{ \eta  }_{ s } \right) &=& K \sum _{ i=1 }^{ N }{ \frac { { { p }_{ i }^{ \mu  } }{ p }_{ i }^{ \nu  } }{ { p }_{ i }^{ \tau  } }  } \frac { 1 }{ { \tau  }_{ 0 } { 2\pi { \sigma  }_{ r }^{ 2 } }  \sqrt { 2\pi { \sigma  }_{ \eta_s }^{ 2 } }  }
\nonumber \\ &&
\hspace{-0.8 in} \times \exp\left[ -\frac { { \left( x-{ x }_{ i } \right)  }^{ 2 }+{ \left( y-{ y }_{ i } \right)  }^{ 2 } }{ 2{ \sigma  }_{ r }^{ 2 } } - \frac { { \left( \eta_s -\eta _{ is } \right)  }^{ 2 } }{ 2{ \sigma  }_{ \eta s }^{ 2 } }  \right],
\label{Tmunu_hydro}
\end{eqnarray}
where ${ p }_{ i }^{ \tau  }={ m }_{ iT }\cosh\left( { Y }_{ i }-{ \eta  }_{ is } \right)$, and
${ p }_{ i }^{ x, y }={ p }_{ x,y i }$, ${ p }_{ i }^{ \eta  }={ m }_{ iT }\sinh\left( { Y }_{ i }-{ \eta }_{ is } \right) /{ \tau  }_{ 0 }$ are the four-momenta of the $i$th parton and $Y_{i}$, $\eta_{is}$, and $m_{iT}$ are the momentum rapidity, the spatial rapidity, and the transverse mass of the $i$th parton, respectively. 
Unless otherwise stated, the smearing parameters are taken as: $\sigma_{ r }$=0.6~fm and $\sigma_{ \eta s }$=0.6 from Refs. \cite{Pang:2012he, Pang:2013pma} where the soft hadron spectra, rapiditydistribution and elliptic flow can be well described. 
The sum index $i$ runs over all produced partons~($N$) in a given nucleus-nucleus collision.
The scale factor $K$ and the initial proper time $\tau_{0}$ are the two free parameters that we adjust to reproduce the experimental measurements of hadron spectra for central Pb+Pb collisions at mid-rapidity \cite{Pang:2012he}.
In the current study, hydrodynamic simulations start at an initial proper time $\tau_{0}= 0.4$~fm/$c$.
After initializing the energy-momentum tensor, we numerically solve the following energy-momentum conservation equation in ($\tau$, $x$, $y$, $\eta_{s}$) coordinates:

 \begin{equation}
\partial_\nu T^{\mu\nu} = 0.
\label{eqTmu}
\end{equation}
The energy-momentum tensor of an ideal fluid in the above equation is defined as,

\begin{equation}
T^{\mu\nu}=(\varepsilon+P)u^\mu u^\nu - Pg^{\mu\nu}.
\end{equation}
where $\varepsilon$, $P$, and $u^{\mu}$ are the energy density, pressure, and fluid four velocity, respectively.

In hydrodynamic model simulations, we use the parametrized equation of state (EoS) s95p-v1~\cite{Huovinen:2009yb} with a cross-over transition between the high-temperature QGP phase and the low-temperature hadronic phase. In this parametrization,  the EoS of the low temperature phase is modeled  by a hadronic resonance gas while the high temperature phase is given by the lattice QCD calculations.
The chemical freeze-out temperature is taken to be 150 MeV. Finally,  hadron momentum distributions are calculated on a hypersurface at a constant kinetic freeze-out temperature~$T_{f}$ using the Cooper-Frye formula~\cite{Cooper:1974mv}.
The kinetic freeze-out temperature for the present study is set to be $T_{f}$=137 MeV.
In our hydrodynamics code, an improved projection method is used to calculate the freeze-out hyper-surface,  which is computationally more efficient than  other conventional methods~\cite{Pang:2012he}.
For the results presented in this study, we have carried out simulations of about 1000 hydrodynamic events for each
centrality bin of heavy-ion collisions.

\subsection{A multiphase transport (AMPT) model}

To compare to the ideal hydrodynamic model simulations, we also  utilize the AMPT model with string melting (version 2.257d) \cite{Lin:2004en}. The initial conditions for the AMPT model are also obtained from the HIJING model~\cite{Wang:1991hta,Gyulassy:1994ew,Wang:2009qb}, which includes minijet partons and excited strings. HIJING uses the Monte Carlo Glauber~(MC Glauber) model (for a review on the Glauber model see Ref.~\cite{Miller:2007ri} ) to simulate nucleon-nucleon collisions  using the Wood-Saxon form for the nuclear density distribution. Each participant nucleon becomes an excited string while each binary nucleon-nucleon collision results in minijet production and further excitation of the participant nucleon-strings. In the current version of the AMPT (with string melting) model, hadrons are produced via string fragmentation of the initial minijets and excited strings which are calculated in the HIJING model from the wounded nucleons and binary collisions, respectively. These hadrons are then converted to their valence quarks and anti-quarks. The space-time evolution of the produced partons is then governed by a parton cascade model ZPC~\cite{Zhang:1997ej} in which only two-parton elastic scattering processes are considered. In the ZPC model, the approximated differential cross sections for parton-parton scattering~($\sigma$) depend on the value of the strong coupling constant $\alpha_{\rm s}$ as well as on the screening mass  ($\mu$) of the partons in the medium. In our current study at the LHC we use two different values for the partonic cross section to investigate the sensitivity of the final results to the interaction strength (and to mimic the sensitivity to the shear viscosity to entropy density ratio) of the system. For one case, we take $\sigma$=1.5 mb following Ref.~\cite{Xu:2011fi}. For the other case, we keep all other input parameters fixed and use a much larger value, $\sigma$=20 mb. At the end of partonic evolution, partons are converted to hadrons through a parton coalescence mechanism. The subsequent hadronic evolution in the AMPT model is carried out through a relativistic transport~(ART) model~\cite{Li:1995pra}.  Other input parameters used in the present version of the AMPT model for Pb+Pb collisions at $\sqrt{s_{\rm NN}}$=2.76 TeV are the strong coupling constant~($\alpha_s$=0.33), and the parameter $a$(0.9), $b$(0.5~GeV$^{2}$) used in the Lund string fragmentation model~(more details about these parameters can be found in Ref.~\cite{Xu:2011fi}). In the current study, about 10000 of Pb+Pb collision events have been performed for each collision centrality bin.

\section{Longitudinal correlations of anisotropic flows}


Using a (3+1)D ideal hydrodynamical model and the AMPT model as briefly described in the above section, we proceed to study
correlations of anisotropic flows (of the same harmonic order) at different rapidities for Pb+Pb collisions at the LHC.
For this purpose, we divide the pseudo-rapidity coverage of $\eta \in (-5.5, 5.5)$ into 11 equal rapidity bins (each has 1 unit of pseudorapidity bin size).  We will use the central pseudo-rapidity value to denote each pseudo-rapidity bin.
We study the correlation between two different pseudo-rapidity bins which we generally label as $A$ and $B$ ($\eta_A \neq \eta_B)$.
The correlation function between the two rapidity bins $A$ and $B$ is defined as,
\begin{eqnarray}
C_n(A, B) =  \frac{\langle \mathbf{Q}_{n}(A) \cdot \mathbf{Q}_{n}^*(B) \rangle }{\sqrt{\langle \mathbf{Q}_{n}(A) \cdot \mathbf{Q}_{n}^*(A) \rangle}\sqrt{ \langle \mathbf{Q}_{n}(B) \cdot \mathbf{Q}_{n}^*(B) \rangle}},
\label{DefCn_hydro}
\end{eqnarray}
where the angular brackets denote the real part of the average over many events.
The vector $\mathbf{Q}_n$ for the $n$th order anisotropic flow in a given pseudo-rapidity bin is defined as,
\begin{eqnarray}
\mathbf{Q}_n \equiv Q_n e^{in\Phi_n}= \frac { 1 }{ N } \sum _{j=1}^{N}{ e^{i n \phi_j}},
\end{eqnarray}
where $\phi_j=\arctan(p_{y j}/p_{x j})$ is the azimuthal angle of the $j$th particle's transverse momentum.

In terms of the magnitudes $Q_n$ and the orientation angles $\Phi_n$ of the flow vectors, the above correlation function becomes
\begin{eqnarray}
C_n(A, B) = \frac{\langle Q_n(A) Q_n(B) e^{i n[\Phi_n(A) - \Phi_n(B)]}\rangle } {\sqrt{\langle Q_n^2(A) \rangle}\sqrt{ \langle Q_n^2(B) \rangle}}.
\end{eqnarray}
 It is clear that when the multiplicity is infinite, or in the continuum limit of hydrodynamic simulations, the above summation becomes an integration over the phase space of each pseudo-rapidity bin. Then the $\mathbf{Q}_n$ vector will be identical to the flow vector $\mathbf{v}_n$, $\mathbf{Q}_n = \mathbf{v}_n$.

For simplicity, we choose $A$ and $B$ rapidity bins in our calculations to be symmetrically located with respect to the mid-rapidity, i.e., $\eta_A = -\eta_B = \eta$. The above correlation function may be simply denoted  as,
\begin{eqnarray}
C_n(\Delta \eta)=C_n(|\eta_A-\eta_B|) = C_n(\eta_A, \eta_B) = C_n(A, B).
\end{eqnarray}
For example, $C_n(\Delta\eta=4)$ represents the correlation between pseudorapidity bins $A$ and $B$ with $\eta_A \in (-2.5, -1.5)$ and $\eta_B \in (1.5, 2.5)$.  It is obvious that the above correlation function becomes unity if hadron spectra in rapidity bins $A$ and $B$ are identical in each event, for example in the case of boost invariance. But in the presence of longitudinal fluctuations, it will deviate from unity. In this article we explore such deviation or decorrelation due to initial-state fluctuations in the longitudinal direction, its dependence on the pseudo-rapidity gap and its sensitivity to transport dynamics of the partonic matter.

\subsection{Longitudinal correlations from hydrodynamic simulations}

\begin{figure}
\begin{center}
\includegraphics[scale=0.55]{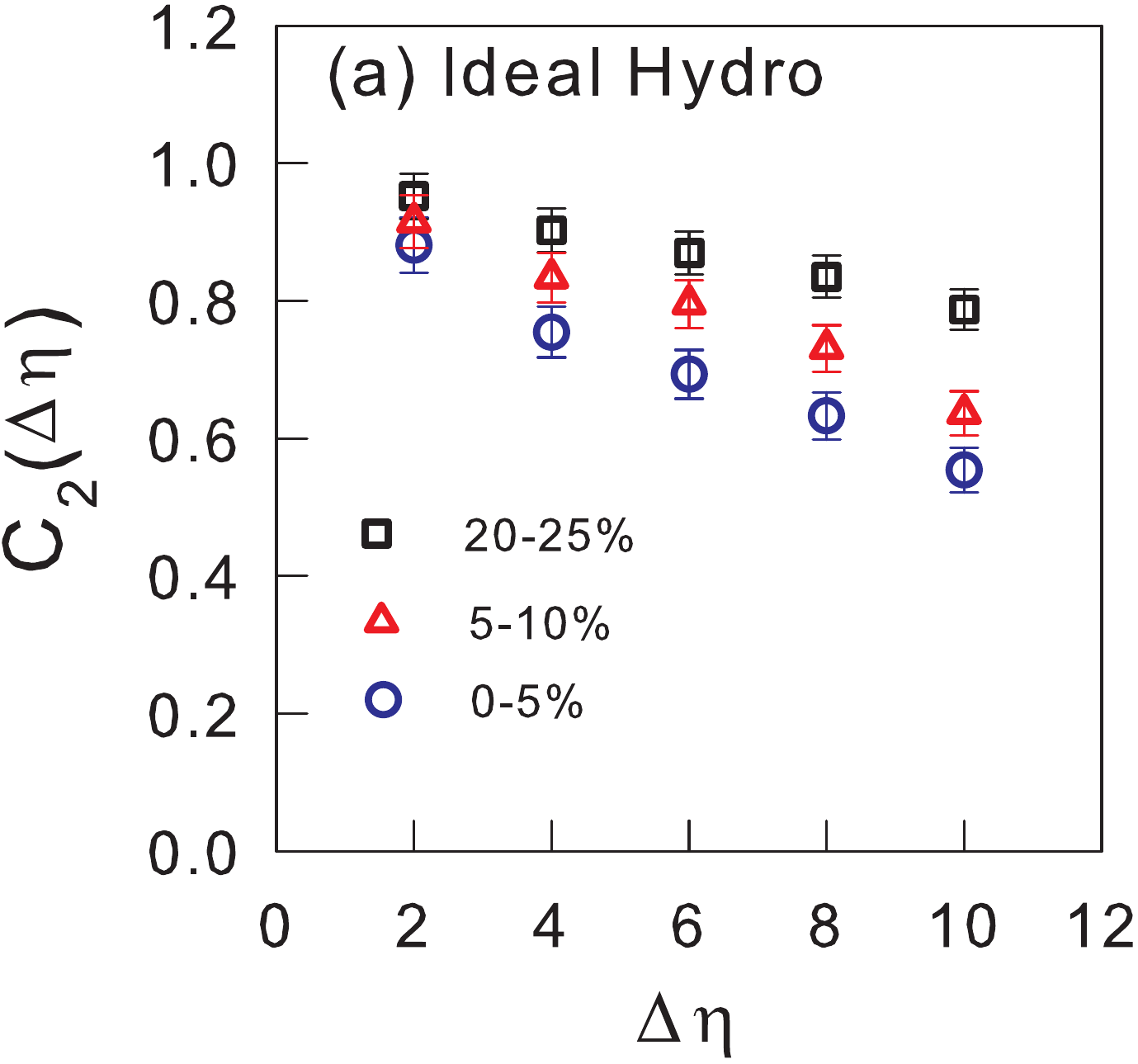}
\includegraphics[scale=0.55]{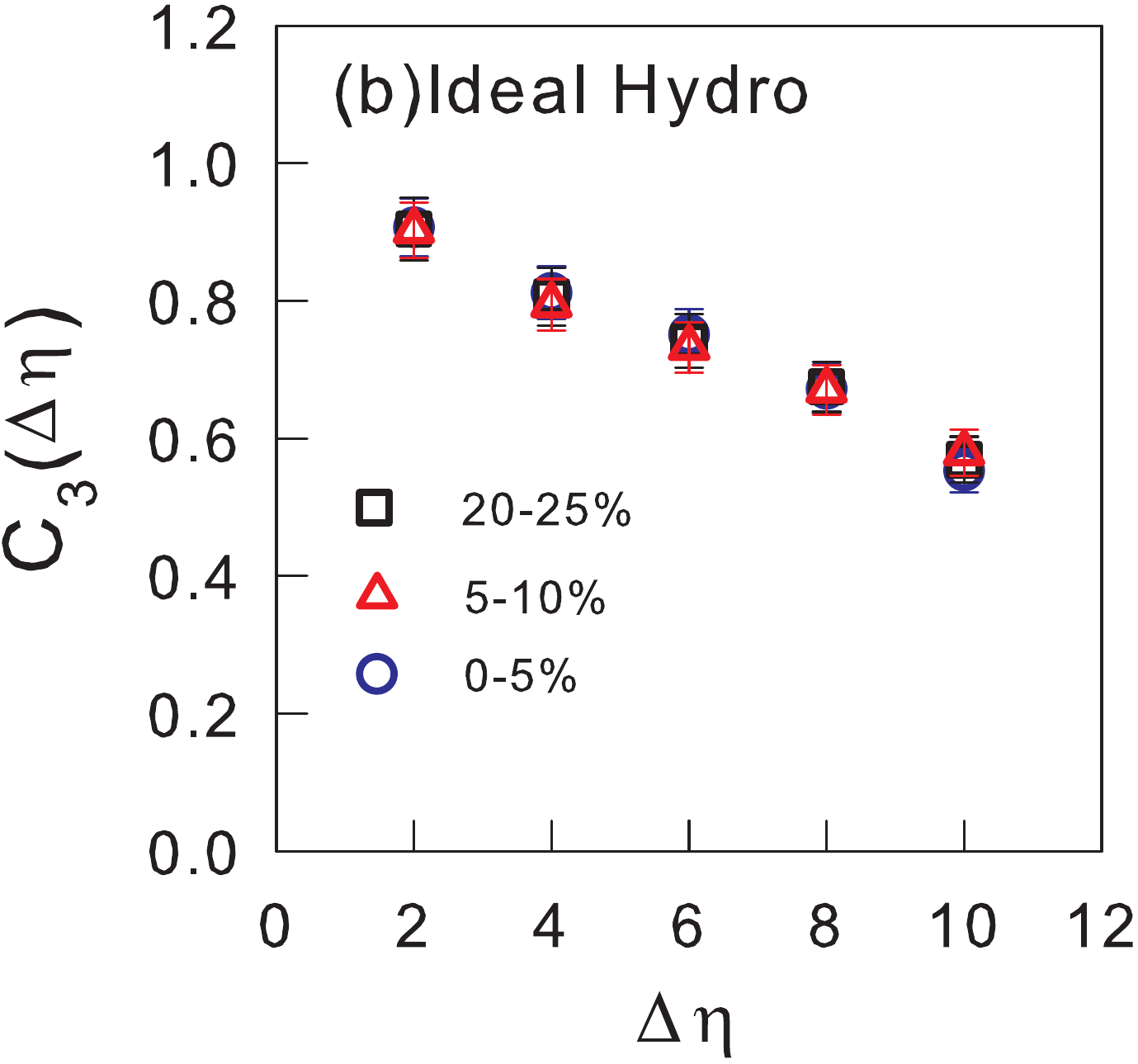}
\caption{(Color online) The longitudinal correlation functions $C_2(\Delta\eta)$ (a) and $C_3(\Delta\eta)$ (b) in
20-25\% (squares), 5-10\% (triangles) and  0-5\% (circles) central  Pb+Pb collisions at $\sqrt{s_{\rm NN}}$=2.76 TeV from ideal hydrodynamic model simulations.}
\label{CentralityHydro}
\end{center}
\end{figure}

\begin{figure}
\begin{center}
\includegraphics[scale=0.55]{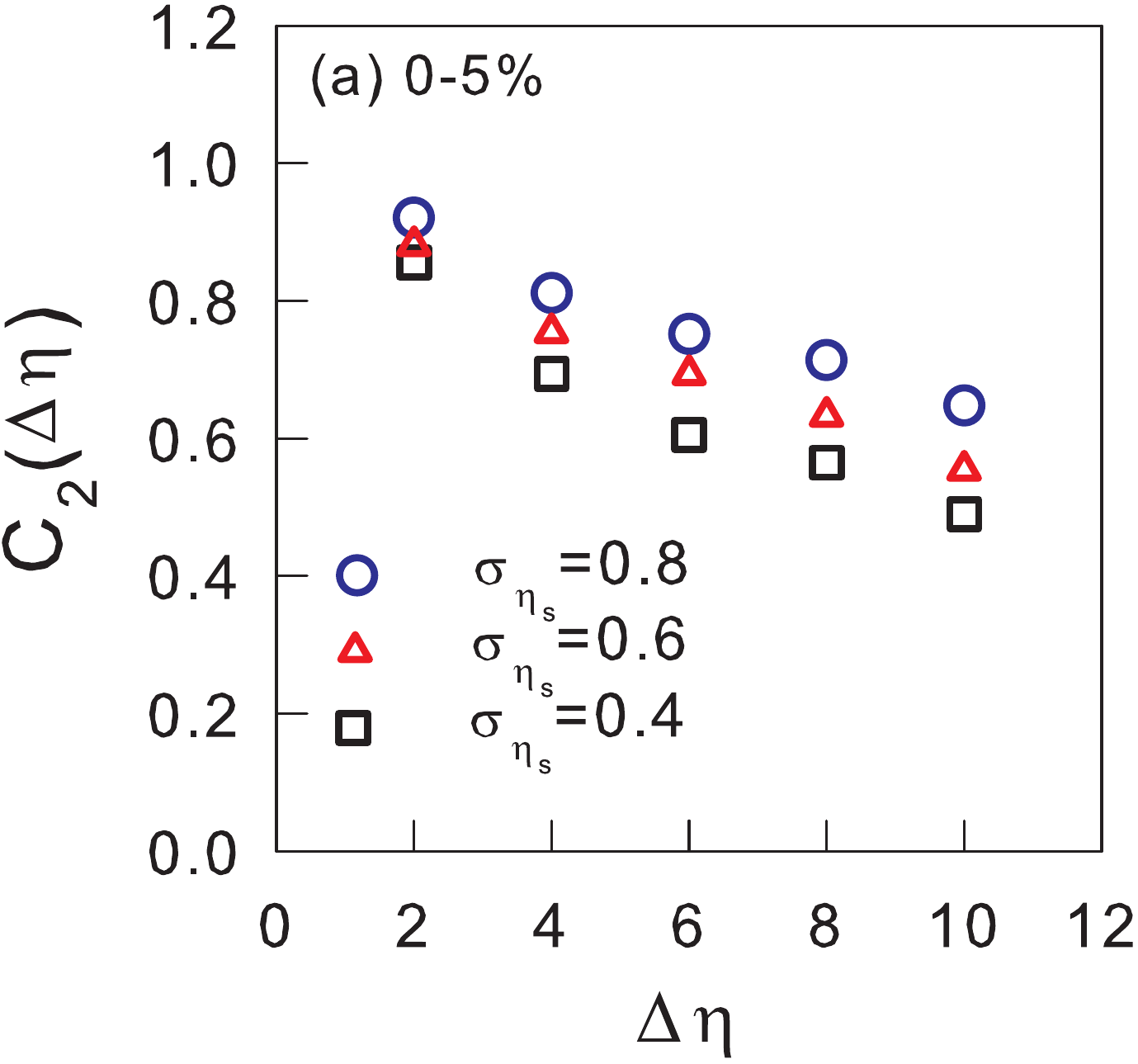}
\includegraphics[scale=0.55]{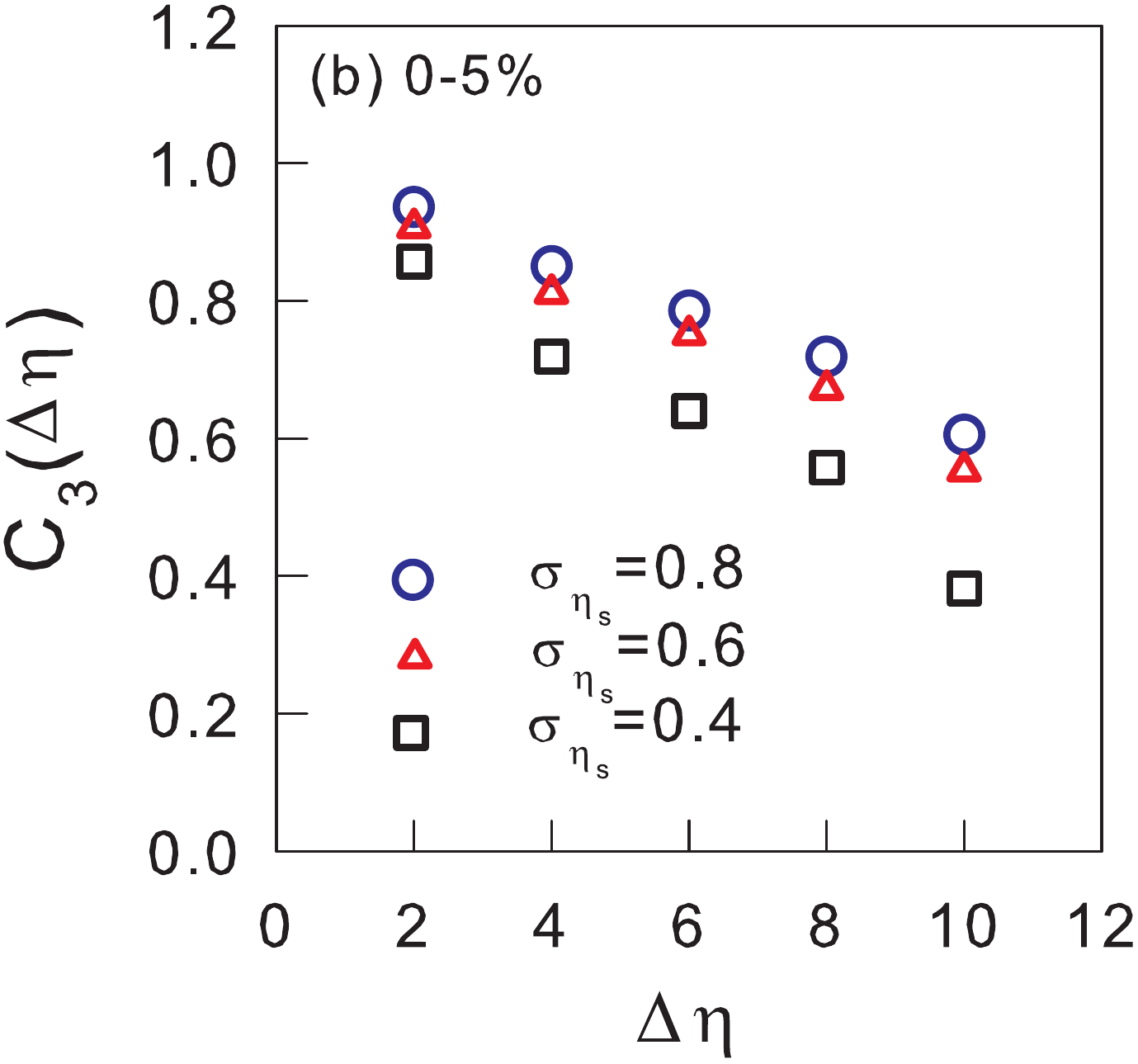}
\caption{(Color online) The dependence of the longitudinal correlation functions $C_2(\Delta\eta)$ (a) and $C_3(\Delta\eta)$ (b) on the smearing parameter $\sigma_{\eta_s}$ in
0-5\% central  Pb+Pb collisions at $\sqrt{s_{\rm NN}}$=2.76 TeV from ideal hydrodynamic model simulations.}
\label{Hydro_vs_sigma}
\end{center}
\end{figure}

We first examine the longitudinal fluctuations and correlations of anisotropic flows from the (3+1)D ideal hydrodynamic model calculations.
In Fig.~\ref{CentralityHydro} we show the correlation functions  $C_2(\Delta\eta)$ [panel (a)] and $C_3(\Delta\eta)$ [panel (b)] for three different centrality bins (squares for 20-25\%, triangles for 5-10\%, and circles for 0-5\%) of Pb+Pb collisions at the LHC. The correlation function will approach unity by definition when the rapidity gap vanishes. As is expected, the anisotropic flows at different pseudorapidities are not perfectly correlated with each other, and the decorrelation becomes more significant as the pseudo-rapidity gap increases. Such decorrelation results mainly from the longitudinal fluctuations in the initial energy density distribution which are translated into the longitudinal fluctuations of the anisotropic flows of the final hadron spectra via hydrodynamic evolution.

From Fig.~\ref{CentralityHydro}, we can also see that the correlation function $C_2$ of the second-order anisotropic flow or elliptic flow shows a strong centrality dependence.  The correlation in semicentral collisions is stronger than that in central collisions. In contrast, the correlation function $C_3$ of the third-order anisotropic flow or triangular flow is independent of centrality.
The zero centrality dependence of $C_3$ may be understood from the fact that the triangular flow $v_3$ arises purely from initial-state fluctuations, which are almost independent of collision geometry. The elliptic flow $v_2$ in the most central heavy-ion collisions is also purely driven by fluctuations in the initial energy density distribution. This explains why the magnitudes of $C_3$ are similar to those of $C_2$ in the most 0-5\% central collisions because both elliptic flow in the most central collisions and triangular flow originate from fluctuations in the initial states. In semi-central collisions, the system develops large elliptic flow $v_2$  due to the collision geometry, which leads to stronger longitudinal correlations of the elliptic flow (larger $C_2$) than in the most central collisions.

The final longitudinal correlation and decorrelation should strongly depend on the initial energy density distribution (and the degree of the fluctuations) in spatial-rapidity.  In hydrodynamic simulations, the degree of longitudinal fluctuations in the initial state is partially controlled by the smearing parameter $\sigma_{\eta_s}$ in the calculation of the initial energy-momentum tensor from the AMPT Monte Carlo simulations in Eq. (\ref{Tmunu_hydro}). One expects that an increase in $\sigma_{\eta_s}$ will lead to flatter initial energy distributions in the rapidity direction, thus the longitudinal decorrelation will be weaker. Such an effect is shown in Fig. \ref{Hydro_vs_sigma}, where the longitudinal correlation functions $C_2(\Delta\eta)$ and $C_3(\Delta\eta)$ are calculated using different values of $\sigma_{\eta_s}$ (0.4, 0.6, and 0.8). One can see that smaller values of $\sigma_{\eta_s}$ in the initial state indeed produce weaker longitudinal correlations in the final state. We should note that the value of  $\sigma_{\eta_s}$ should also be constrained by the final hadron rapidity distribution, two-hadron correlations, anisotropic flows and their rapidity dependence as has been done in Refs.~\cite{Pang:2012he, Pang:2013pma}.

\subsection{Longitudinal correlations from AMPT simulations}

As we have illustrated in the ideal hydrodynamic model simulations, the decorrelation of anisotropic flows of final-state hadrons with large pseudo-rapidity gaps arises from the longitudinal fluctuations in the initial energy density distribution through hydrodynamic evolution. One can in principle investigate the sensitivity of this decorrelation to the transport properties of the dense medium by introducing viscous corrections to the ideal hydrodynamical model. Before such a (3+1)-D viscous hydrodynamical model with full fluctuating initial conditions becomes available, we employ the AMPT model for this purpose in this study which uses the same initial conditions as those in the ideal hydrodynamic model simulations.

\begin{figure}
\begin{center}
\includegraphics[scale=0.55]{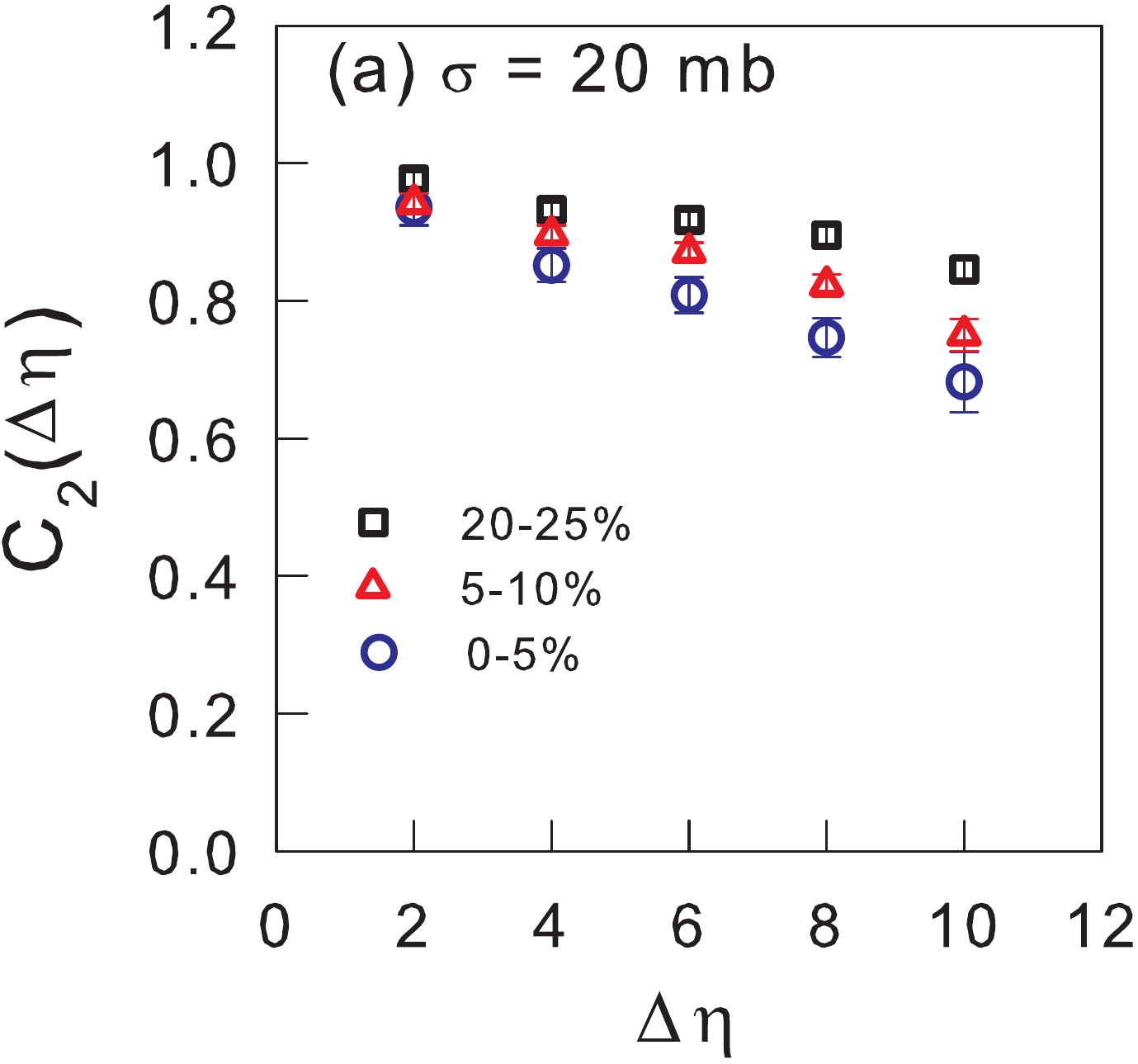}
\includegraphics[scale=0.55]{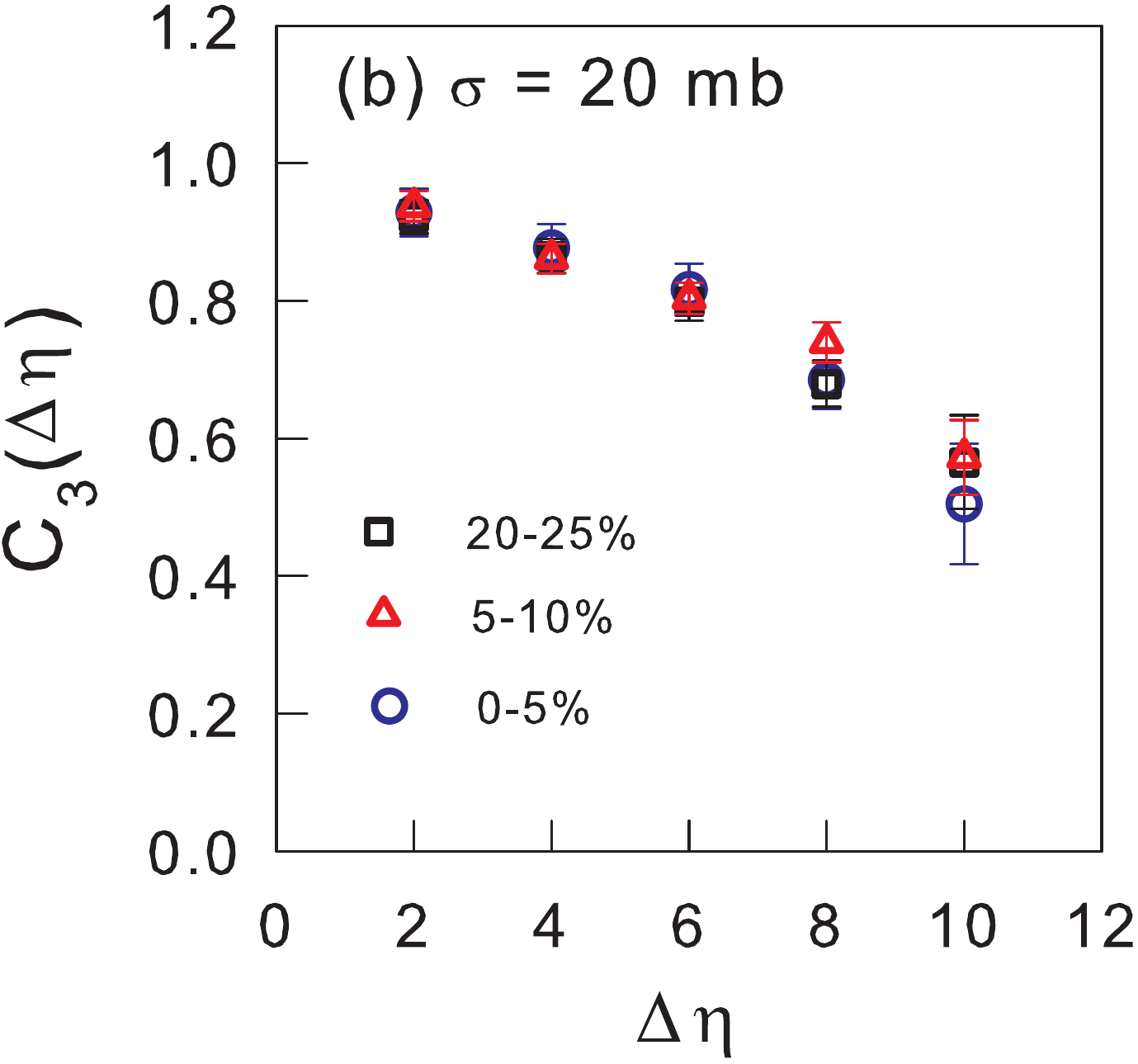}
\caption{(Color online)  The same as Fig.~\ref{CentralityHydro} except from the AMPT model simulations with a partonic cross section $\sigma=20$ mb.}
\label{CentralityAMPT}
\end{center}
\end{figure}

One difference between the hydrodynamic model calculations and the AMPT model simulations is that each AMPT Monte Carlo event has finite hadron multiplicity which will lead to additional statistical fluctuations. We follow the common practice and use a sub-event method to correct the correlations for the effect due to finite multiplicity \cite{Poskanzer:1998yz,Ollitrault:1993ba,Bhalerao:2013ina}. To adopt the sub-event method for the calculation of longitudinal correlations with a large pseudo-rapidity gap,  we further randomly divide hadrons in each of the two pseudo-rapidity bins ($A$ or $B$) into two groups or sub-events with equal multiplicity. Thus there is a  total of four subevents which can be labeled as ($A_1$, $A_2$) and ($B_1$, $B_2$) in each event. To correct for the effect of finite hadron multiplicity in the AMPT simulations, we construct the following correlation function between two pseudo-rapidity bins $A$ and $B$:
\begin{eqnarray}
C_n(A, B) = \frac{\frac{1}{4} \sum_{i,j=1,2} \langle \mathbf{Q}_n(A_i) \cdot \mathbf{Q}_n^*(B_j) \rangle }{\sqrt{\langle \mathbf{Q}_n(A_1) \cdot \mathbf{Q}_n^*(A_2) \rangle}\sqrt{ \langle \mathbf{Q}_n(B_1) \cdot \mathbf{Q}_n^*(B_2) \rangle}},
\label{DefCn_ampt}
\end{eqnarray}
where $\mathbf{Q}_n(A_i)$ and $\mathbf{Q}_n(B_i)$ are the $\mathbf{Q}_n$ vectors for $A_i$ and $B_i$  sub-event.

\begin{figure}
\begin{center}
\includegraphics[scale=0.55]{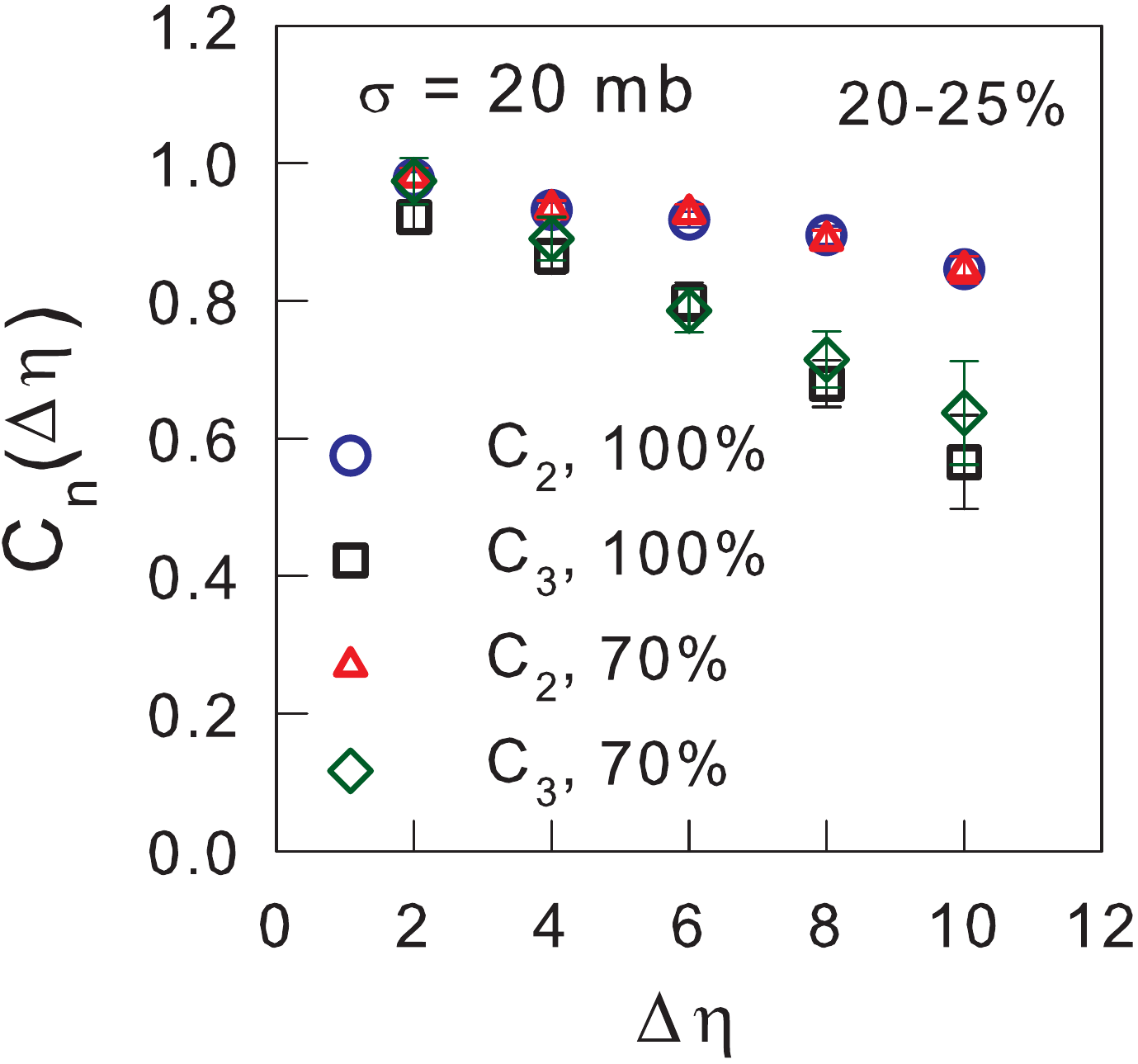}
\caption{(Color online) Correlation functions $C_2(\Delta\eta)$ (circles and triangles)  and $C_3(\Delta\eta)$ (squares and rhombuses) calculated with 100\% ( circles and squares) and 70\% of charged hadron multiplicity (triangles and rhombus) in 20-25$\%$ Pb+Pb collisions at $\sqrt{s_{\rm NN}}$=2.76 TeV from the AMPT model simulations with a partonic cross section $\sigma=20$ mb.}
\label{MultiplicityEffect}
\end{center}
\end{figure}

\begin{figure}
\begin{center}
\includegraphics[scale=0.55]{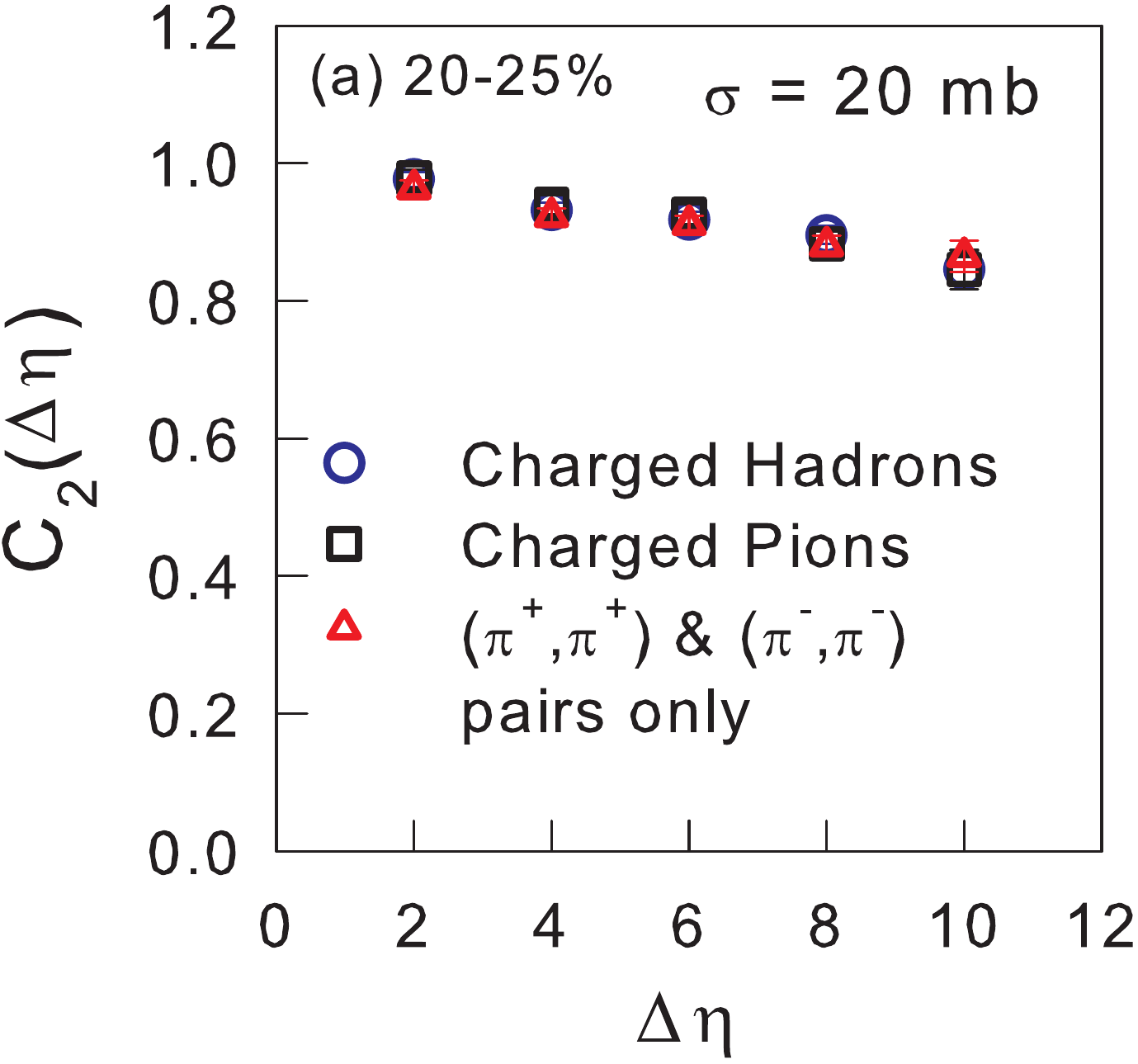}
\includegraphics[scale=0.55]{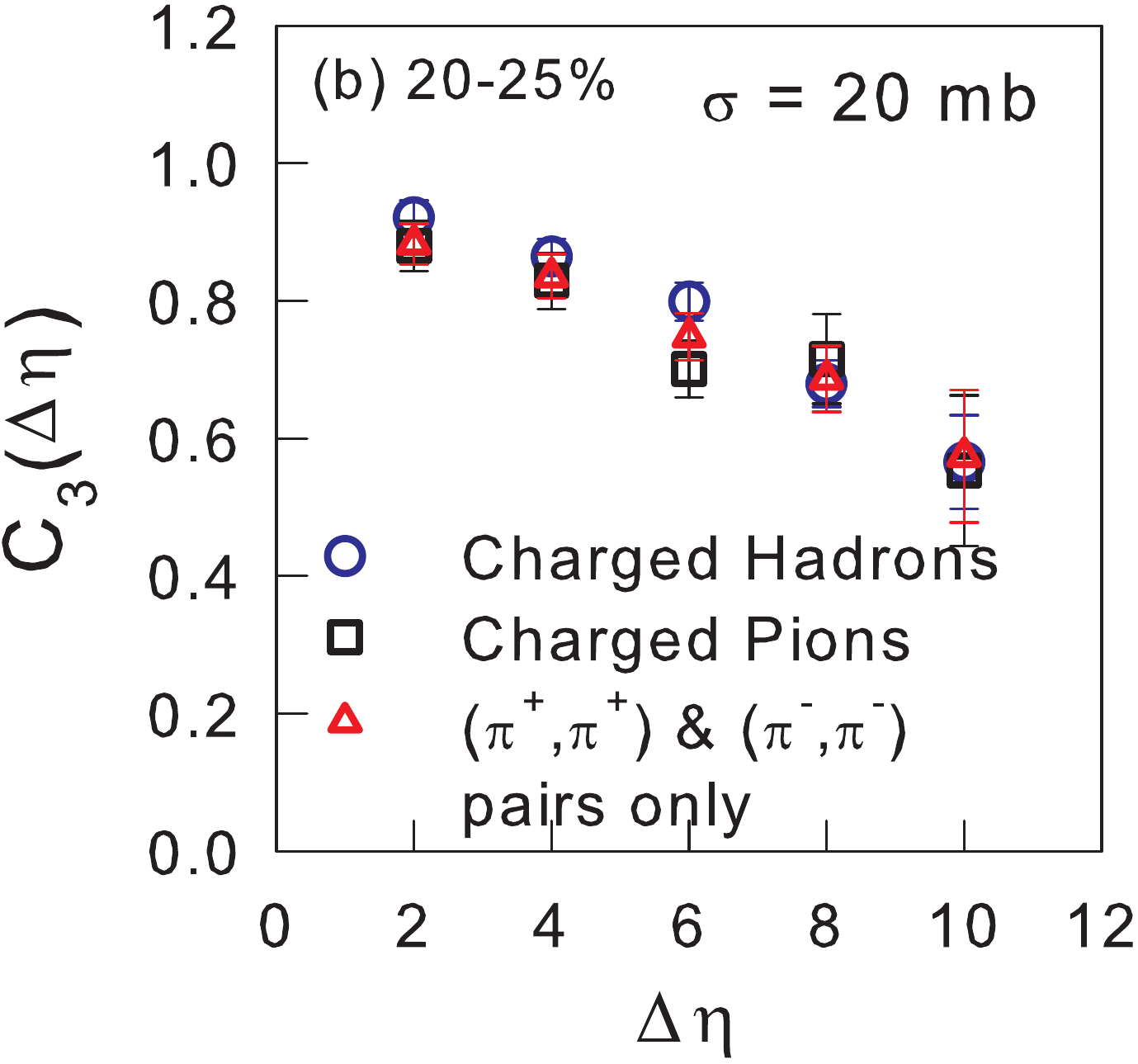}
\caption{(Color online) Correlation functions  $C_2(\Delta\eta)$ (a) and $C_3(\Delta\eta)$ (b) calculated
with all charged hadrons (circles), all charged pions  (squares) and pions with the same signs of charge (triangles)
 in 20-25$\%$ Pb+Pb collisions at $\sqrt{s_{\rm NN}}$=2.76 TeV  from the AMPT model simulations with a partonic cross section $\sigma=20$ mb.}
\label{chargeEffect}
\end{center}
\end{figure}

In Fig.~\ref{CentralityAMPT} we show the longitudinal correlation functions $C_2(\Delta\eta)$ [panel (a)] and $C_3(\Delta\eta)$ [panel (b)] from the AMPT model simulations of Pb+Pb collisions at the LHC. The parton cross section is taken to be $\sigma=20$~mb. Again, the results for three different collision centralities for both $C_2$ and $C_3$ are compared (squares for 20-25\%, triangles for 5-10\%, and circles for 0-5\% centrality). One can see that results from the AMPT model are generally similar to the results from the ideal hydrodynamic model simulations as shown in Fig.~\ref{CentralityHydro}.
Correlations between anisotropic flows at different pseudorapidities decrease with increasing pseudo-rapidity gaps.
Strong centrality dependence is observed for $C_2$ but not for $C_3$, and the magnitudes of $C_3$ are close to those of $C_2$ in the most 0-5\% central collisions. We have also checked that results for $\sigma=1.5$~mb show similar properties.

As we have mentioned, the above correlation functions between two different rapidity bins in Eq.~(\ref{DefCn_ampt}) are constructed to minimize the effect of finite multiplicity in the AMPT Monte Carlo simulations. To check the effectiveness of such a subevent method to minimize the effect of finite multiplicity, we calculate the correlation functions again from the same simulations but with only $70\%$ of randomly selected charged particles in each event and compare them to the results using all charged particles as shown in Fig.~\ref{MultiplicityEffect}, where circles and triangles are for $C_2$ and squares and rhombuses are for $C_3$. We can see that the results for different multiplicities  agree quite well within statistical errors for both $C_2$ and $C_3$. This indicates that the correlation functions calculated with the sub-event method contain little contributions from statistical fluctuations due to finite multiplicity in each event.

In the sub-event method that we use to calculate the correlation functions, the two sub-events ($A_1$ and $A_2$, or $B_1$ and $B_2$) in the same rapidity bin might introduce some non-flow contributions (such as resonance decays) to the calculated correlation functions.
To investigate the effect of non-flow contributions, we calculate the correlation functions $C_n({\Delta\eta)}$ for three different selections of final hadrons: all charge hadrons (circles), all charged pions (squares), and pions with the same signs of charge (triangles), i.e., ($\pi^{+},\pi^{+}$) and ($\pi^{-},\pi^{-}$) as shown in Fig. \ref{chargeEffect} for $C_2$ [panel (a)] and $C_3$ [panel (b)].
We can see that the longitudinal correlations of anisotropic flows in all three different cases agree with each other within the statistical errors,
suggesting that there are negligible non-flow effects such as that from resonance decays in the longitudinal correlations that we study here. There are still possible nonflow contributions from mini-jets. These nonflow effects from minijets, however, contribute only about 2\% to the overall dihadron correlations \cite{Pang:2013pma}. One therefore expects similar small contributions to the anisotropic flow correlations.


\subsection{Dependence on the evolution dynamics}

\begin{figure}
\begin{center}
\includegraphics[scale=0.55]{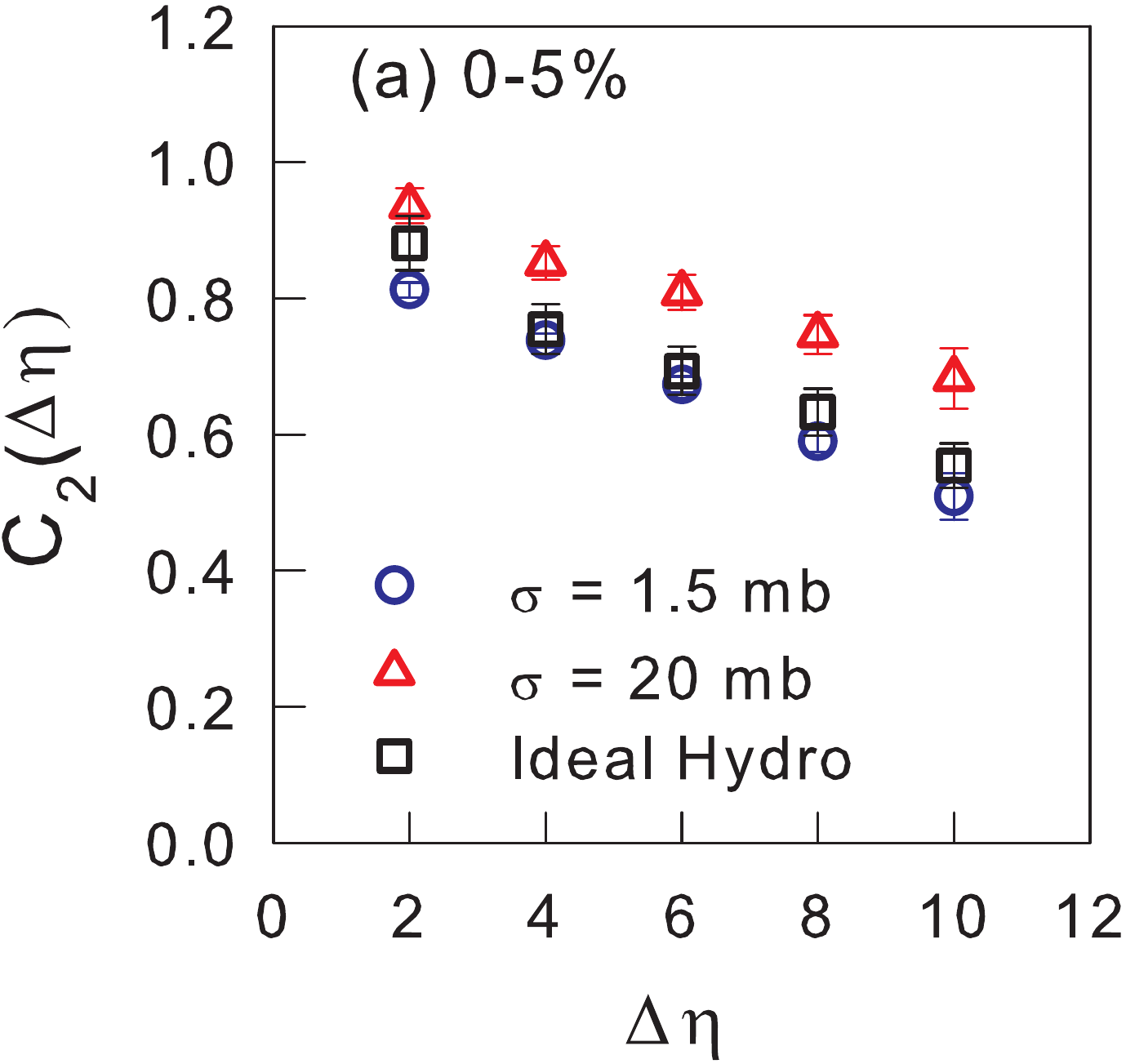}
\includegraphics[scale=0.55]{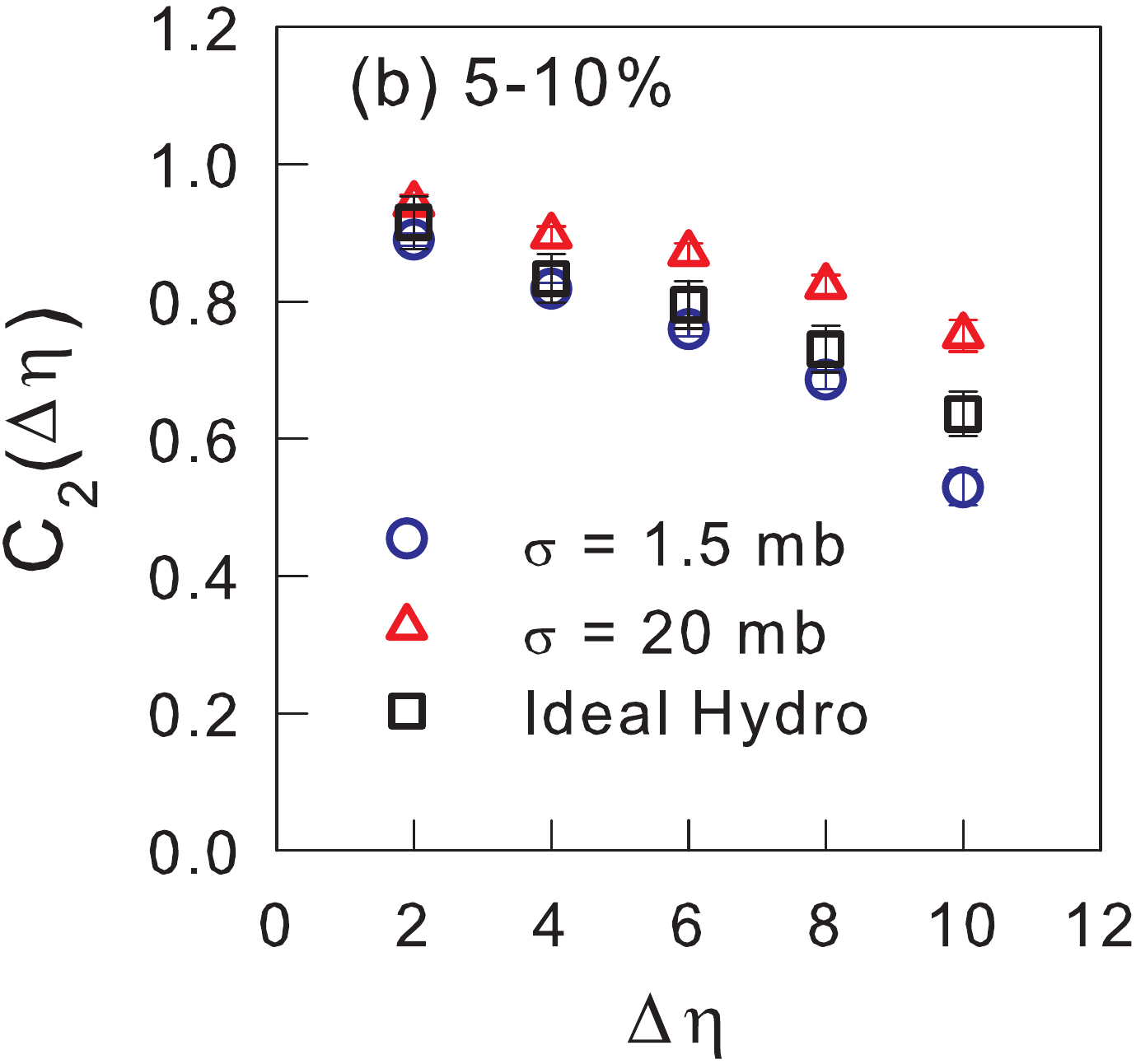}
\includegraphics[scale=0.55]{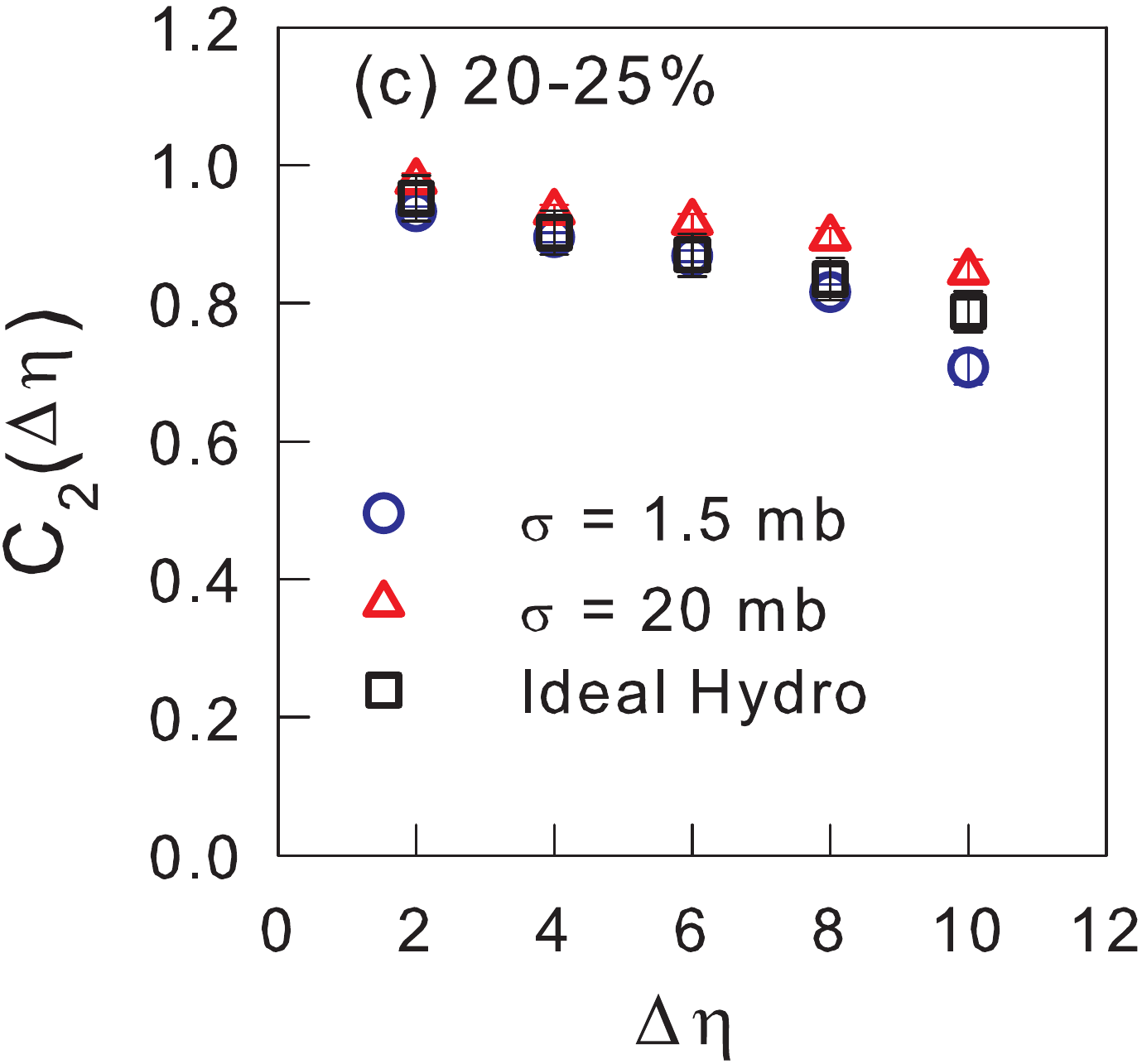}
\caption{(Color online) The correlation function $C_2(\Delta\eta)$ from ideal hydrodynamic (squares) and the AMPT model simulations of
0-5\% (a), 5-10\% (b) and 20-25\% (c) central Pb+Pb collisions at $\sqrt{s_{\rm NN}}=2.76$ TeV with two different values of the parton cross section $\sigma =1.5$ mb (circles)  and 20 mb (triangles).}
\label{Ep2LHC0t5-c2}
\end{center}
\end{figure}

\begin{figure}
\begin{center}
\includegraphics[scale=0.55]{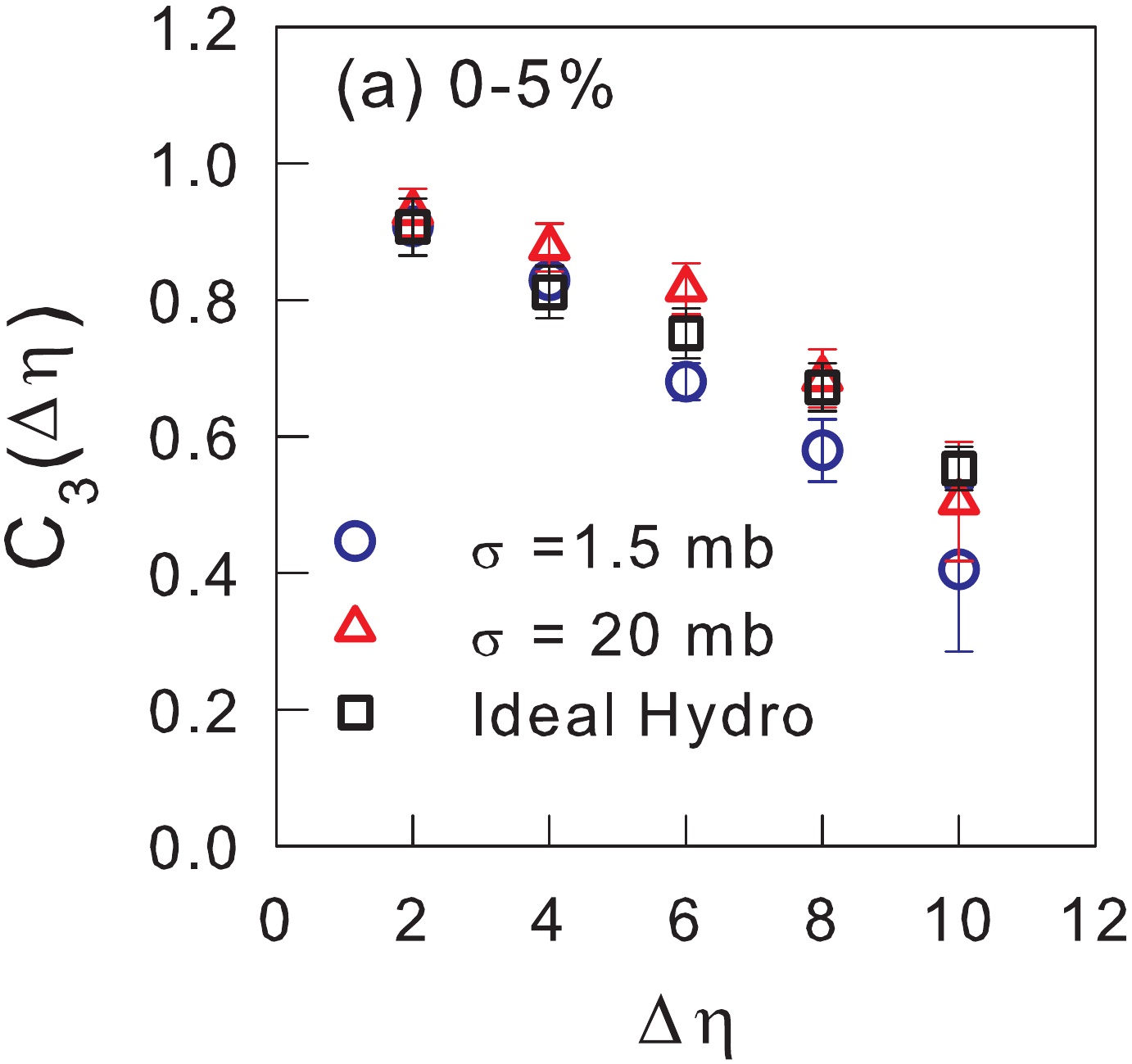}
\includegraphics[scale=0.55]{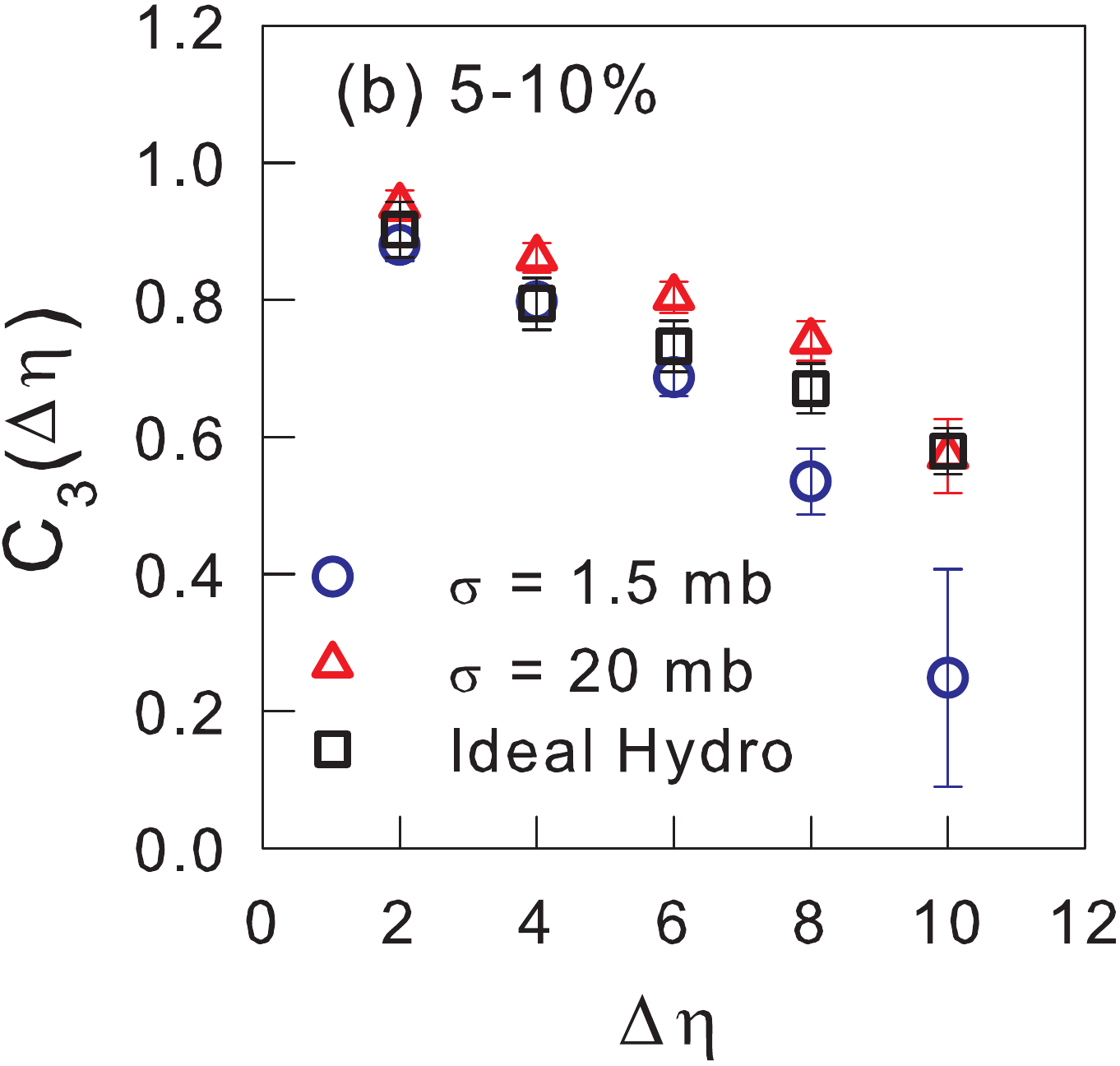}
\includegraphics[scale=0.55]{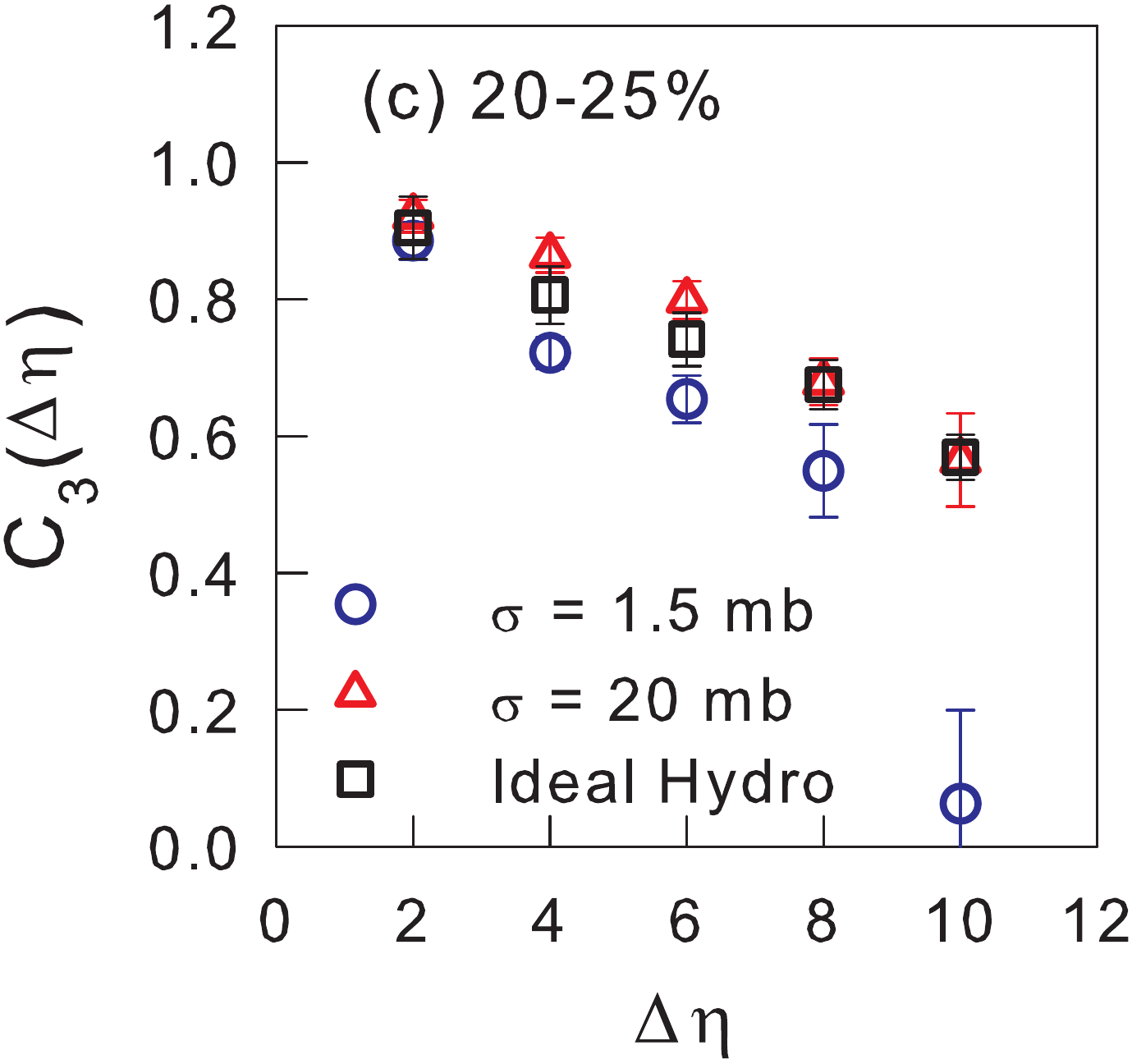}
\caption{(Color online) The same as Fig.~\ref{Ep2LHC0t5-c2} except for the correlation function $C_3(\Delta\eta)$.}
\label{Ep2LHC0t5-c3}
\end{center}
\end{figure}

As we have seen in the last two subsections,  both the ideal hydrodynamical model and the AMPT model give similar longitudinal correlations ($C_2$ and $C_3$) and their dependence on the pseudo-rapidity gaps and collision centrality. To explore their sensitivities to the evolution dynamics, we compare in this subsection the AMPT model results with different values of parton cross sections and the results from the ideal hydrodynamic model simulations, which corresponds to the strong coupling limit of a transport model. Without a viscous  (3+1)D hydrodynamical model with full fluctuating initial conditions (in both transverse and longitudinal direction) at hand now, the variation with the parton cross section in this study provides some hints on the effect of shear viscosity on the longitudinal correlations.

Shown in Figs.~\ref{Ep2LHC0t5-c2} and \ref{Ep2LHC0t5-c3} are the correlation functions $C_2(\Delta\eta)$ and $C_3(\Delta\eta)$
from  the AMPT model simulations of Pb+Pb collisions at the LHC with two different values of parton cross sections $\sigma=1.5$ mb (circles) and 20 mb (triangles) for centrality class 0-5\% [panels (a)], 5-10\% [panels (b)] and 20-25\% [panels (c)]. It is clear that a larger parton cross section or a stronger interaction strength in the AMPT model tends to increase the longitudinal correlations as it also increases the collectivity of the system. Because the effective shear viscosity to entropy density ratio in the AMPT model is inversely proportional to the parton cross section, the AMPT model results also indicate a possible dependence of the final-state longitudinal correlations on the shear viscosity to entropy ratio of the partonic matter. An increased parton cross section also increases the final hadron multiplicity in the AMPT model. However, this increased multiplicity from the same initial energy density distributions should not affect the normalized correlations $C_n(\Delta\eta)$. A variation in the parton cross section in the AMPT model, however, also leads to a change in the effective EoS of the partonic matter, which could influence the strength of the longitudinal correlations.

We also compare the AMPT results with the  ideal hydrodynamic model calculations of the longitudinal correlations (squares) in Figs.~\ref{Ep2LHC0t5-c2} and \ref{Ep2LHC0t5-c3}, which are between the AMPT results with two different values of the parton cross section.
This is a little counterintuitive because one would generally expect that the ideal hydrodynamical model should be the strong coupling limit when the parton cross section is infinitely large.  Such an intuitive expectation, however, is complicated by many other differences between the AMPT model and the ideal hydrodynamical model we use. For example, the EoS of the partonic matter in the AMPT model should be different from what we use in the ideal hydrodynamical model \cite{Zhang:2008zzk}. Although we have the same initial conditions for both the hydrodynamical model and the AMPT model, a Gaussian smearing is used to obtain the initial energy density distribution for the later hydrodynamic evolution. In addition, the ideal hydrodynamical model and the AMPT model produce different values of anisotropic flows $v_n$. These and other differences might be responsible for the deviation from the naive expectation on the comparisons between the AMPT and the ideal hydrodynamic model results. The use of viscous (3+1)-D hydrodynamical models in the future would be able to clarify this question and show the effect of shear viscosity on the longitudinal correlations.

\begin{figure}
\begin{center}
\includegraphics[scale=0.55]{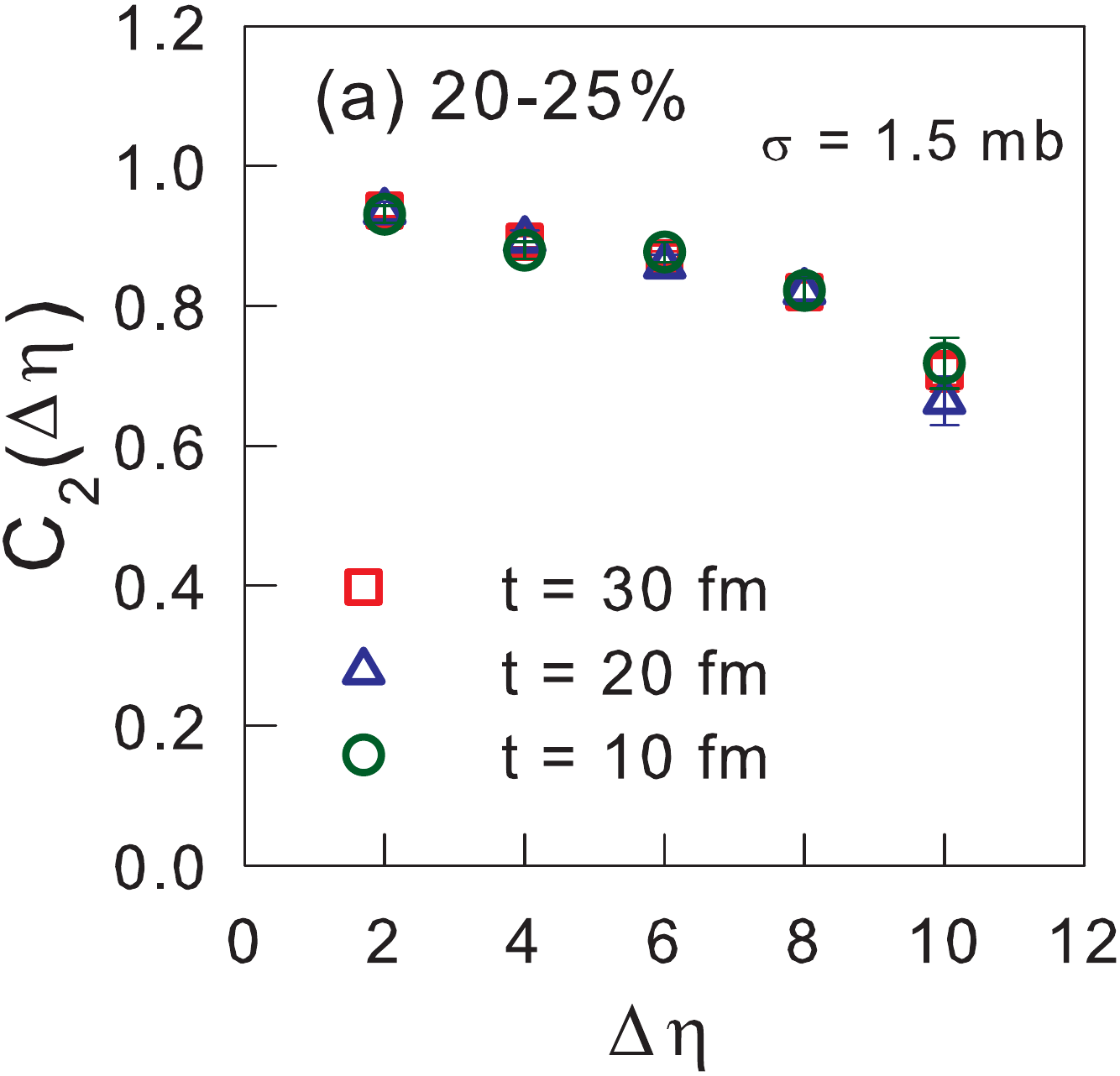}
\includegraphics[scale=0.55]{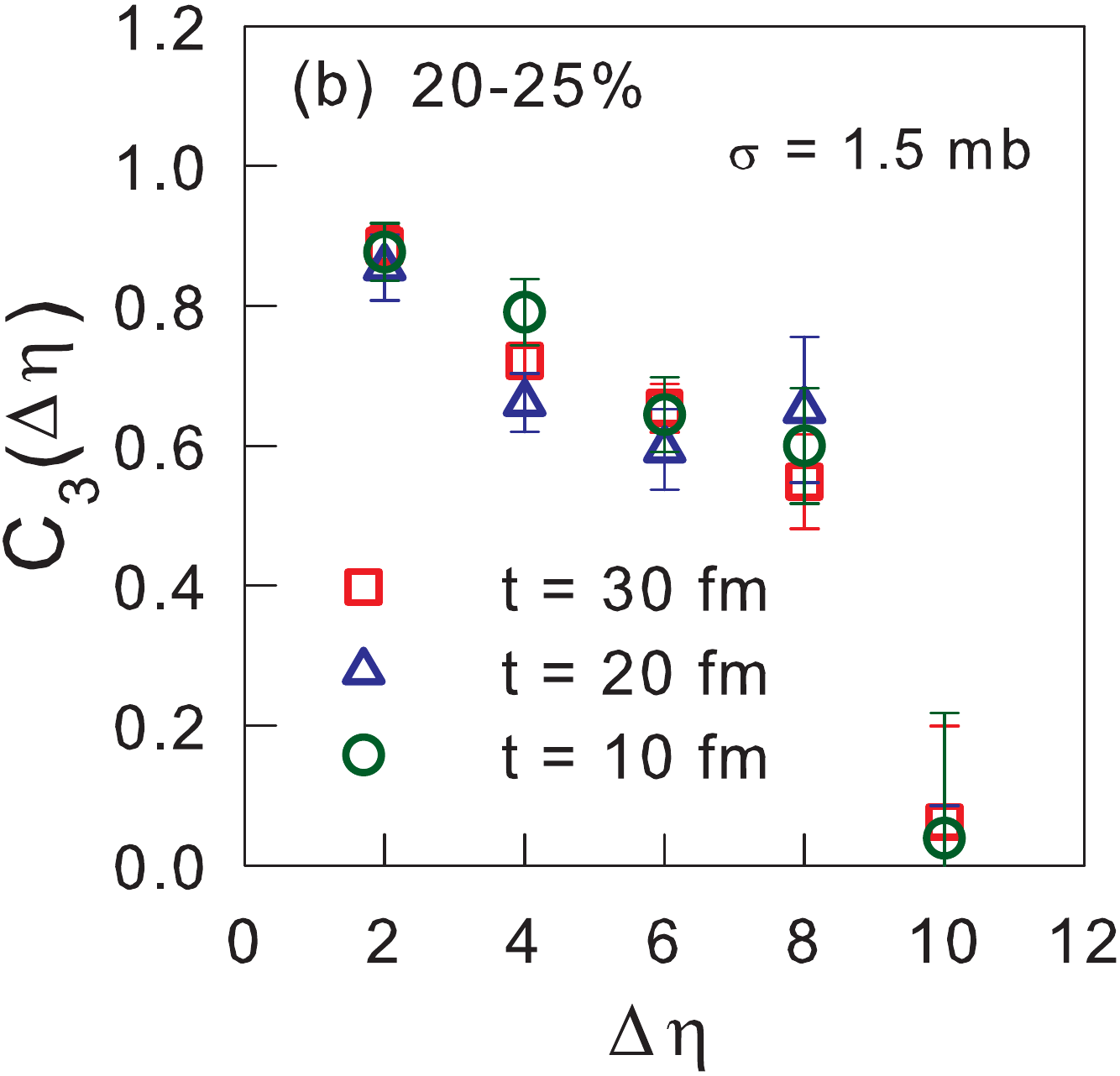}
\caption{(Color online) Correlation functions $C_2(\Delta\eta)$ (a) and $C_3(\Delta\eta)$ (b) for different hadronic evolution times ($t=$10, 20 and 30 fm). The partonic cross section in the AMPT model is taken to be $\sigma $ = 1.5 mb. }
\label{DifferentTime}
\end{center}
\end{figure}


Because the final-state correlations between anisotropic flows at different pseudo-rapidities as shown above depend on the evolution dynamics of the partonic matter,  it is thus interesting to investigate whether they are also influenced by the hadronic interaction in the late stage of the fireball evolution.
For this purpose, we vary the hadronic evolution time in the AMPT model while keeping all other conditions fixed. Shown in Fig.~\ref{DifferentTime} are the longitudinal correlations $C_2(\Delta\eta)$ [panel (a)] and $C_3(\Delta\eta)$ [panel (b)]  from the AMPT model simulations of 20-25\% central Pb+Pb collisions at the LHC with a partonic cross section $\sigma$ = 1.5 mb at different evolution times $t=30$, 20, and 10 fm, when the hadronic phase dominates in the interacting matter.  We can see that the correlation functions remain unchanged for different hadronic evolution time, indicating that the long-range longitudinal correlations have already been built up during the partonic phase.

\subsection{Twist versus fluctuations}

\begin{figure}
\begin{center}
\includegraphics[scale=0.38]{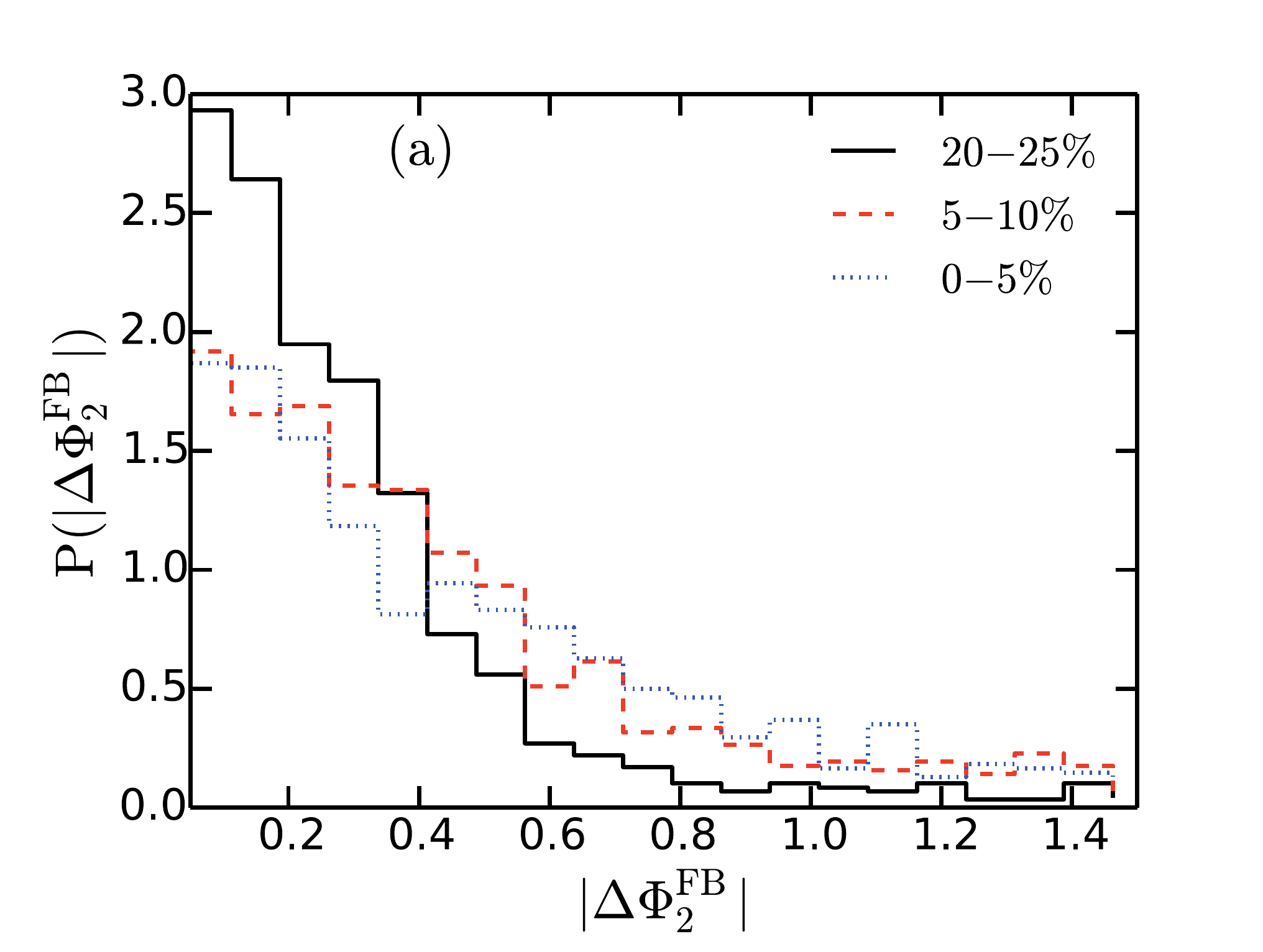}
\includegraphics[scale=0.38]{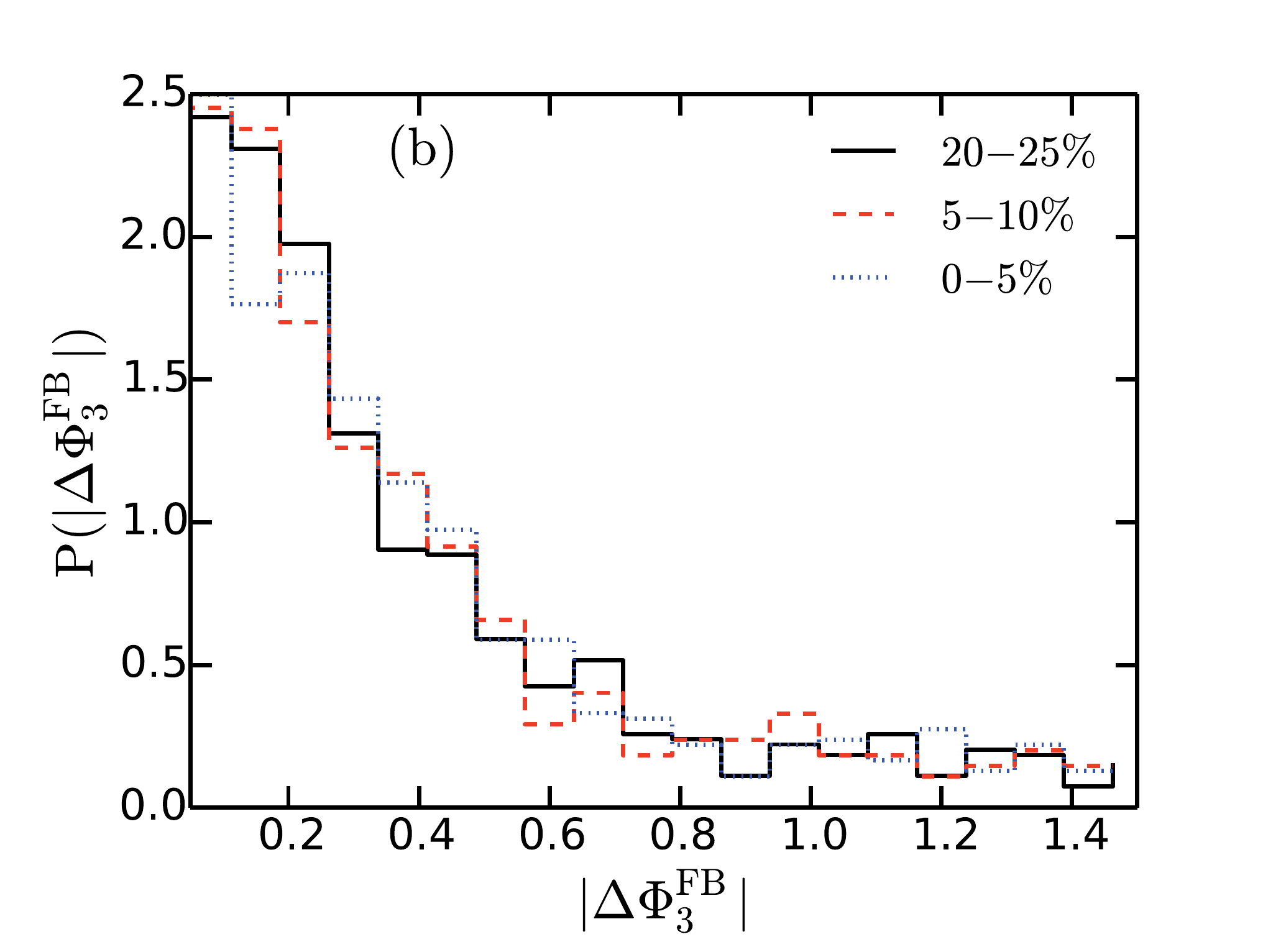}
\caption{(Color online) Event distributions in the twist angles $|\Delta\Phi_2^{\rm FB}|$ (a) and $|\Delta\Phi_3^{\rm FB}|$ (b) from hydrodynamic simulations of Pb+Pb collisions at $\sqrt{s_{\rm NN}}=2.76$ TeV with different centralities. }
\label{twistangle}
\end{center}
\end{figure}

In a recent study within the AMPT model \cite{Jia:2014ysa} the longitudinal distribution of event plane angles is found to have a systematic twist or rotation between forward and backward directions. Such a twist in the AMPT model is caused by different contributions to the initial particle production from projectile and target participant nucleons and the difference in the initial participant angles in the forward and backward rapidity regions \cite{Bozek:2010vz}. It should also be partially responsible for the long-range longitudinal decorrelations observed in this study.
The twist in each event can be characterized by the twist angle or the flow angle difference $\Delta\Phi_n^{\rm FB}=\Phi_n(F)-\Phi_n(B)$ in the forward and backward pseudo-rapidity regions. Here, $F$ ($B$) represents the last (first) of the 11 pseudo-rapidity bins in the interval $\eta \in (-5.5, 5.5)$ in this study.  Shown in the Fig.~\ref{twistangle} are the event distributions in the twist angles $|\Delta\Phi_2^{\rm FB}|$ (upper panel) and $|\Delta\Phi_3^{\rm FB}|$ (lower panel) from our hydrodynamic simulations of Pb+Pb collisions with different centralities at the LHC. With initial conditions from the AMPT(HIJING) model, the distributions in the twist angles are quite broad. They should lead to a systematic decorrelation of the anisotropic flows along the longitudinal direction (pseudo-rapidity). The distributions in the twist angle of the third harmonics are independent of the centrality, reflecting the fluctuating nature of the triangular flow. The distributions in  $|\Delta\Phi_2^{\rm FB}|$, however, become narrower in more peripheral collisions because the elliptic flow is mainly driven by the geometrical shape of the overlap region of the initial particle production. This is partially responsible for the centrality dependence of the longitudinal correlation of the elliptic flow as shown in Fig.~\ref{CentralityHydro}.

\begin{figure}
\begin{center}
\includegraphics[scale=0.38]{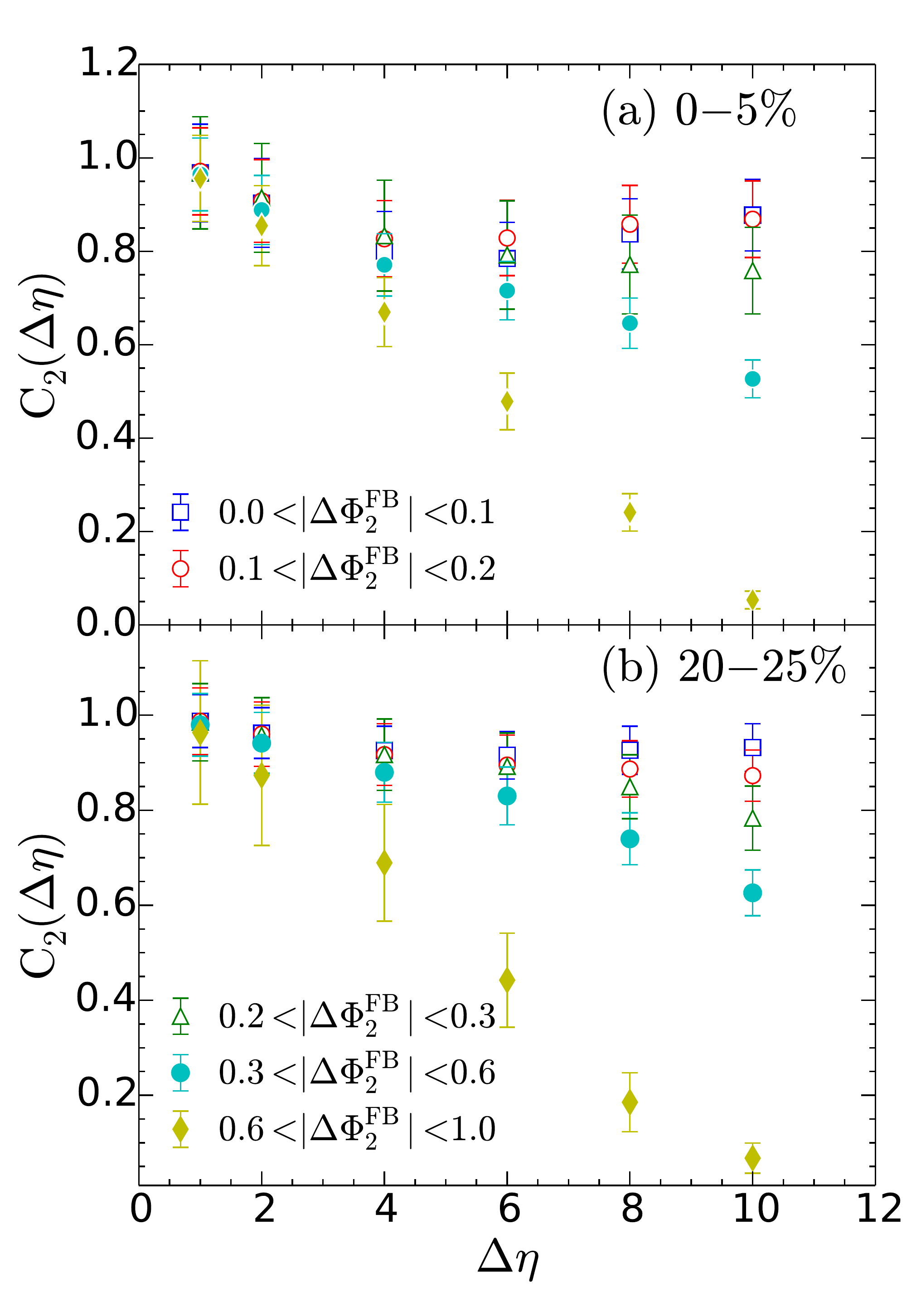}
\caption{(Color online) Correlation functions $C_2(\Delta\eta)$ for different twist angles $\Delta\Phi_2^{\rm FB}$ from hydrodynamic simulations of 0-5\% (a) and 20-25\%  (b) central Pb+Pb collisions at $\sqrt{s_{\rm NN}}=2.76$ TeV. }
\label{twist2}
\end{center}
\end{figure}

\begin{figure}
\begin{center}
\includegraphics[scale=0.38]{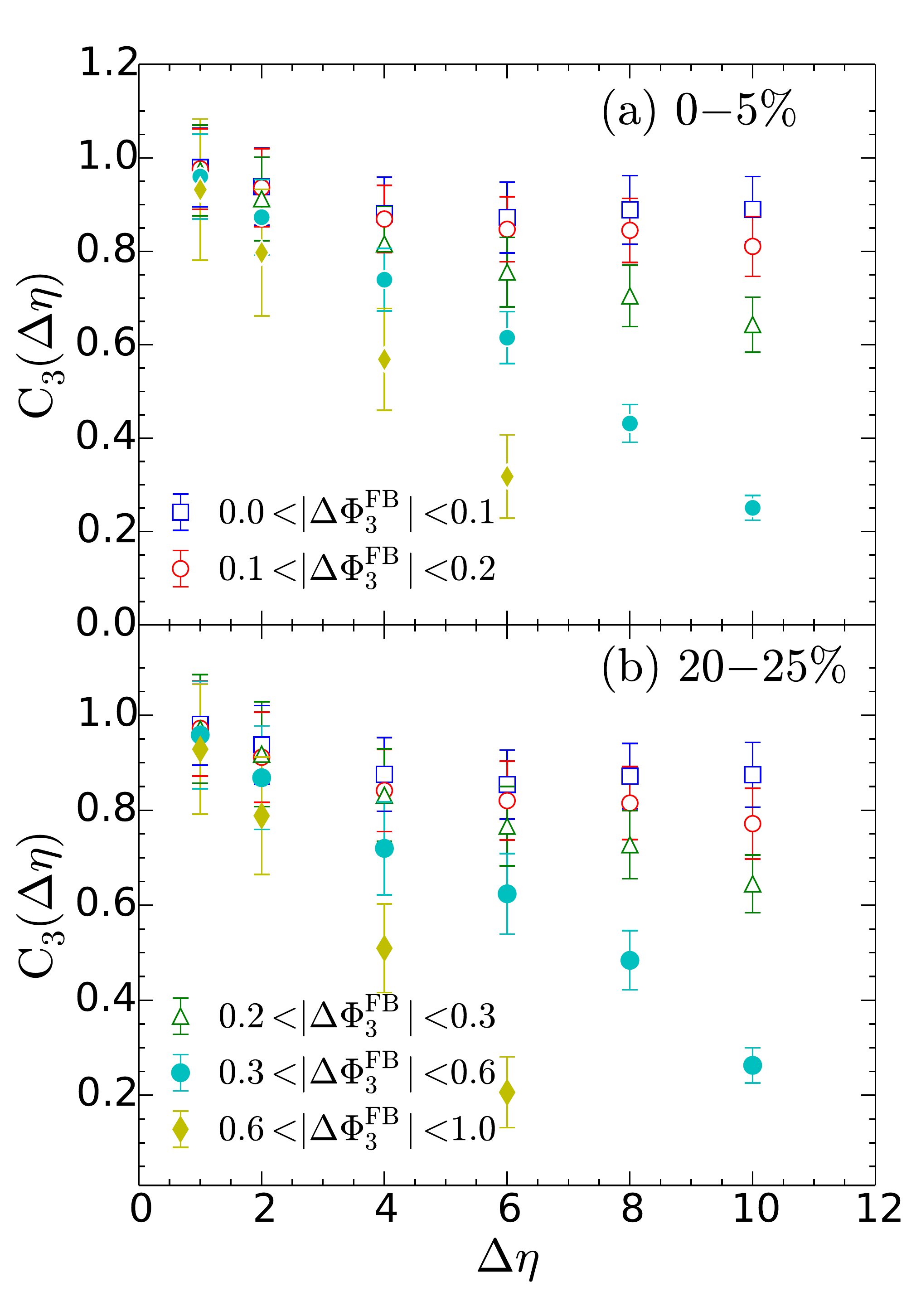}
\caption{(Color online) Correlation functions $C_3(\Delta\eta)$ for different twist angles $\Delta\Phi_3^{\rm FB}$ from hydrodynamic simulations of 0-5\% (a) and 20-25\% (b) central Pb+Pb collisions at $\sqrt{s_{\rm NN}}=2.76$ TeV. }
\label{twist3}
\end{center}
\end{figure}

To separate any additional long-range decorrelation of anisotropic flows on top of the twist, one can impose restrictions on the twist angles.
For small twist angles $\Delta\Phi_n^{\rm FB}$, one expects additional long-range decorrelation on top of that caused by the twist.
Shown in Figs.~\ref{twist2}  and \ref{twist3} are the longitudinal correlation functions $C_n(\Delta\eta)$ $(n=2, 3)$ for different twist angles $|\Delta\Phi_n^{\rm FB}|$ from hydrodynamic simulations of 0-5\% [panel (a)] and 20-25\% [panel (b)] central Pb+Pb collisions at $\sqrt{s_{\rm NN}}=2.76$ TeV. For large values of the twist angle $|\Delta\Phi_n^{\rm FB}|$ of the $n$-th harmonics, the longitudinal correlation functions $C_n(\Delta\eta)$ are significantly smaller than 1, especially for large pseudo-rapidity gaps. They should approach the limit,
\begin{equation}
C_n^{\rm FB}=\frac{\langle v_n^{\rm F} v_n^{\rm B}\rangle}{\sqrt{\langle {v_n^{\rm F}}^2\rangle \langle {v_n^{\rm B}}^2\rangle }}\cos(n\Delta\Phi_n^{\rm FB})
\end{equation}
when the two pseudo-rapidity bins are close to the most forward and backward regions, where $v_n^{\rm F, B}$ are the amplitudes of the anisotropic flows in the forward and backward regions. This limit even becomes negative for $|\Delta\Phi_n^{\rm FB}|\ge \pi/2n$ as we see in our hydrodynamic simulations.  Note that the above forward-backward correlation of elliptic flows was also proposed to separate the non-flow effect from jet production \cite{Liao:2009ni,Han:2011ys}. The effect of the initial twist will make such separation difficult.

As the values of the twist angle become close to zero, the longitudinal correlation functions approach the limiting functions which can be attributed to pure longitudinal fluctuations. The limiting longitudinal correlations have a similar form for both elliptic and triangular flow. The magnitude of the limiting correlation functions for the elliptic flow depends on the centrality of the collisions, similar to the case without restrictions on the twist angle. It increases for noncentral events because of larger geometrically driven elliptic flow. The magnitude of the limiting longitudinal correlation function for $C_3(\Delta\eta)$ is, however, independent of the centrality because the triangular flow is driven by pure transverse fluctuations in all centralities. We have also calculated $C_3(\Delta\eta)$ for events with different twist angles of the second harmonics $|\Delta\Phi_2^{\rm FB}|$ and $C_2(\Delta\eta)$ for different values of $|\Delta\Phi_3^{\rm FB}|$. These correlations functions do not depend on the twist angles, because the triangular flow is driven purely by the transverse fluctuations and is uncorrelated to the elliptic flow.

Because the last bin of the pseudorapidity gap coincides with our defined forward-backward gap, the values of the limiting longitudinal correlation functions reflect the correlation of the amplitudes of the anisotropic flows in the forward and backward region,
\begin{equation}
c_n^{\rm FB}=\frac{\langle v_n^{\rm F} v_n^{\rm B}\rangle}{\sqrt{\langle {v_n^{\rm F}}^2 \rangle \langle {v_n^{\rm B}}^2\rangle }}.
\end{equation}
As we see from Figs.~\ref{twist2}  and \ref{twist3}, the above correlations of amplitudes of  the anisotropic flows is about $c_n^{\rm FB}\approx 0.9$ for purely fluctuation-driven triangular flow and elliptic flow in the most central collisions. One can calculate the above amplitude correlations for other values of the pseudo-rapidity gap by restricting the flow angle difference $|\Phi_n(\eta)-\Phi_n(-\eta)|\rightarrow 0$.  One can then extract fluctuations of the flow angles on top of the twist due to the difference in participant angles of the projectile and target nucleons.

\section{SUMMARY AND DISCUSSIONS}
\label{sec:summary}

We have studied the longitudinal fluctuations in the initial states of heavy-ion collisions and their effects on the final-state long-range longitudinal correlations. In particular, we have calculated the correlations between anisotropic flows at different pseudo-rapidities in Pb+Pb collisions at the LHC from (3+1)D ideal relativistic hydrodynamic and the AMPT model simulations. For both the hydrodynamic and the AMPT model calculations, we use identical initial conditions from the HIJING simulations which include both transverse and longitudinal fluctuations.

The anisotropic flows at different pseudo-rapidities are found to become partially decorrelated due to longitudinal fluctuations in the initial energy density distribution. The degree of such decorrelations increases when the pseudo-rapidity gap becomes larger.
The correlations for the second-order anisotropic flows are found to depend on the centrality of the collisions, becoming stronger in semi-central collisions than in the most central collisions, due to the geometry-driven elliptic flow in semi-central collisions. The correlations for the third-order anisotropic flow are almost independent of centrality because the triangular flow is purely driven by initial-state fluctuations.

By restricting the difference in flow angles in the most forward and backward pseudo-rapidity regions or in the twist angles, we have further shown the longitudinal de-correlaton in anisotropic flows are caused by longitudinal fluctuations with a twist structure as well as as additional random fluctuations on top of a twist.

Within the AMPT model, we found that the longitudinal correlations depend on the interaction strength of the partonic matter, but not on the hadronic evolution. These results point to the possibility that the longitudinal fluctuations and correlations may be utilized to probe the transport properties, such as the shear viscosity to entropy density ratio, of the hot and dense partonic matter created in heavy-ion collisions.

The observed decorrelation of anisotropic flows at different pseudo-rapidities can also provide one explanation for the difference between the elliptic flows calculated from (2+1)D and (3+1)D hydrodynamical models with fluctuating initial conditions \cite{Pang:2012he}. In these calculations, one determines the event-plane angle $\Psi_n^{\rm EP}$ from the hadron spectrum in the forward and backward pseudorapidity regions. The anisotropic flow coefficients for hadrons in the central pseudorapidity region are then calculated with respect to the event-plane angle in each event. In the (3+1)D hydrodynamic calculations with both transverse and longitudinal fluctuations, the event-plane angles in the central, forward, and backward regions are different due to the longitudinal fluctuations. The resulting elliptic flow should be smaller than in the case of (2+1)D hydrodynamic simulations with no longitudinal fluctuations. It is therefore important to take into account the longitudinal fluctuations for a more accurate extraction of the shear viscosity to entropy density ratio from comparisons of the experimental data and full (3+1)D viscous hydrodynamic simulations.

Future investigations can also include the use of different models for the initial condition to study the effect of different longitudinal fluctuation profiles, and the use of (3+1)D viscous hydrodynamical models to study the effect of shear viscosity on final-state longitudinal correlations. One can also study the longitudinal correlations of anisotropic flows in small systems in p+p and p+A collisions. These studies will be helpful for a more comprehensive understanding of initial-state fluctuations and for more precise determination of the transport properties of the produced fireball in ultra-relativistic nucleus-nucleus collisions.

\section*{Acknowledgments}

We thank J. Jia, W. Li, M. Luzum and B. Mohanty for helpful discussions. This work is supported in part by the Natural Science Foundation of China under Grants No. 11221504, No. 11375072 and No. 11175232, by the Chinese Ministry of Science and Technology under Grant No. 2014DFG02050, and by the Director, Office of Energy Research, Office of High Energy and Nuclear Physics, Division of Nuclear Physics, of the U.S. Department of Energy under Contract No. DE- AC02-05CH11231 and within the framework of the JET Collaboration.


\bibliography{hydroampt_refs}

\begin{thebibliography}{66}
\expandafter\ifx\csname natexlab\endcsname\relax\def\natexlab#1{#1}\fi
\expandafter\ifx\csname bibnamefont\endcsname\relax
  \def\bibnamefont#1{#1}\fi
\expandafter\ifx\csname bibfnamefont\endcsname\relax
  \def\bibfnamefont#1{#1}\fi
\expandafter\ifx\csname citenamefont\endcsname\relax
  \def\citenamefont#1{#1}\fi
\expandafter\ifx\csname url\endcsname\relax
  \def\url#1{\texttt{#1}}\fi
\expandafter\ifx\csname urlprefix\endcsname\relax\def\urlprefix{URL }\fi
\providecommand{\bibinfo}[2]{#2}
\providecommand{\eprint}[2][]{\url{#2}}

\bibitem[{\citenamefont{Adams et~al.}(2004)}]{Adams:2003zg}
\bibinfo{author}{\bibfnamefont{J.}~\bibnamefont{Adams}} \bibnamefont{et~al.}
  (\bibinfo{collaboration}{STAR}), \bibinfo{journal}{Phys. Rev. Lett.}
  \textbf{\bibinfo{volume}{92}}, \bibinfo{pages}{062301}
  (\bibinfo{year}{2004}), \eprint{nucl-ex/0310029}.

\bibitem[{\citenamefont{Aamodt et~al.}(2010)}]{Aamodt:2010pa}
\bibinfo{author}{\bibfnamefont{K.}~\bibnamefont{Aamodt}} \bibnamefont{et~al.}
  (\bibinfo{collaboration}{The ALICE Collaboration}),
  \bibinfo{journal}{Phys.Rev.Lett.} \textbf{\bibinfo{volume}{105}},
  \bibinfo{pages}{252302} (\bibinfo{year}{2010}), \eprint{1011.3914}.

\bibitem[{\citenamefont{Aad et~al.}(2012{\natexlab{a}})}]{ATLAS:2011ah}
\bibinfo{author}{\bibfnamefont{G.}~\bibnamefont{Aad}} \bibnamefont{et~al.}
  (\bibinfo{collaboration}{ATLAS Collaboration}), \bibinfo{journal}{Phys.Lett.}
  \textbf{\bibinfo{volume}{B707}}, \bibinfo{pages}{330}
  (\bibinfo{year}{2012}{\natexlab{a}}), \eprint{1108.6018}.

\bibitem[{\citenamefont{Chatrchyan
  et~al.}(2013{\natexlab{a}})}]{Chatrchyan:2012ta}
\bibinfo{author}{\bibfnamefont{S.}~\bibnamefont{Chatrchyan}}
  \bibnamefont{et~al.} (\bibinfo{collaboration}{CMS Collaboration}),
  \bibinfo{journal}{Phys.Rev.} \textbf{\bibinfo{volume}{C87}},
  \bibinfo{pages}{014902} (\bibinfo{year}{2013}{\natexlab{a}}),
  \eprint{1204.1409}.

\bibitem[{\citenamefont{Heinz and Snellings}(2013)}]{Heinz:2013th}
\bibinfo{author}{\bibfnamefont{U.}~\bibnamefont{Heinz}} \bibnamefont{and}
  \bibinfo{author}{\bibfnamefont{R.}~\bibnamefont{Snellings}},
  \bibinfo{journal}{Ann.Rev.Nucl.Part.Sci.} \textbf{\bibinfo{volume}{63}},
  \bibinfo{pages}{123} (\bibinfo{year}{2013}), \eprint{1301.2826}.

\bibitem[{\citenamefont{Gale et~al.}(2013{\natexlab{a}})\citenamefont{Gale,
  Jeon, and Schenke}}]{Gale:2013da}
\bibinfo{author}{\bibfnamefont{C.}~\bibnamefont{Gale}},
  \bibinfo{author}{\bibfnamefont{S.}~\bibnamefont{Jeon}}, \bibnamefont{and}
  \bibinfo{author}{\bibfnamefont{B.}~\bibnamefont{Schenke}},
  \bibinfo{journal}{Int.J.Mod.Phys.} \textbf{\bibinfo{volume}{A28}},
  \bibinfo{pages}{1340011} (\bibinfo{year}{2013}{\natexlab{a}}),
  \eprint{1301.5893}.

\bibitem[{\citenamefont{Song}(2013)}]{Song:2013gia}
\bibinfo{author}{\bibfnamefont{H.}~\bibnamefont{Song}} (\bibinfo{year}{2013}),
  \eprint{arXiv 1401.0079}.

\bibitem[{\citenamefont{Romatschke}(2010)}]{Romatschke:2009im}
\bibinfo{author}{\bibfnamefont{P.}~\bibnamefont{Romatschke}},
  \bibinfo{journal}{Int.J.Mod.Phys.} \textbf{\bibinfo{volume}{E19}},
  \bibinfo{pages}{1} (\bibinfo{year}{2010}), \eprint{0902.3663}.

\bibitem[{\citenamefont{Huovinen}(2013)}]{Huovinen:2013wma}
\bibinfo{author}{\bibfnamefont{P.}~\bibnamefont{Huovinen}},
  \bibinfo{journal}{Int.J.Mod.Phys.} \textbf{\bibinfo{volume}{E22}},
  \bibinfo{pages}{1330029} (\bibinfo{year}{2013}), \eprint{1311.1849}.

\bibitem[{\citenamefont{Luzum and Petersen}(2014)}]{Luzum:2013yya}
\bibinfo{author}{\bibfnamefont{M.}~\bibnamefont{Luzum}} \bibnamefont{and}
  \bibinfo{author}{\bibfnamefont{H.}~\bibnamefont{Petersen}},
  \bibinfo{journal}{J.Phys.} \textbf{\bibinfo{volume}{G41}},
  \bibinfo{pages}{063102} (\bibinfo{year}{2014}), \eprint{1312.5503}.

\bibitem[{\citenamefont{Alver and Roland}(2010)}]{Alver:2010gr}
\bibinfo{author}{\bibfnamefont{B.}~\bibnamefont{Alver}} \bibnamefont{and}
  \bibinfo{author}{\bibfnamefont{G.}~\bibnamefont{Roland}},
  \bibinfo{journal}{Phys. Rev.} \textbf{\bibinfo{volume}{C81}},
  \bibinfo{pages}{054905} (\bibinfo{year}{2010}), \eprint{1003.0194}.

\bibitem[{\citenamefont{Alver et~al.}(2010)\citenamefont{Alver, Gombeaud,
  Luzum, and Ollitrault}}]{Alver:2010dn}
\bibinfo{author}{\bibfnamefont{B.~H.} \bibnamefont{Alver}},
  \bibinfo{author}{\bibfnamefont{C.}~\bibnamefont{Gombeaud}},
  \bibinfo{author}{\bibfnamefont{M.}~\bibnamefont{Luzum}}, \bibnamefont{and}
  \bibinfo{author}{\bibfnamefont{J.-Y.} \bibnamefont{Ollitrault}}
  (\bibinfo{year}{2010}), \eprint{1007.5469}.

\bibitem[{\citenamefont{Petersen et~al.}(2010)\citenamefont{Petersen, Qin,
  Bass, and Muller}}]{Petersen:2010cw}
\bibinfo{author}{\bibfnamefont{H.}~\bibnamefont{Petersen}},
  \bibinfo{author}{\bibfnamefont{G.-Y.} \bibnamefont{Qin}},
  \bibinfo{author}{\bibfnamefont{S.~A.} \bibnamefont{Bass}}, \bibnamefont{and}
  \bibinfo{author}{\bibfnamefont{B.}~\bibnamefont{Muller}},
  \bibinfo{journal}{Phys.Rev.} \textbf{\bibinfo{volume}{C82}},
  \bibinfo{pages}{041901} (\bibinfo{year}{2010}), \eprint{1008.0625}.

\bibitem[{\citenamefont{Staig and Shuryak}(2011)}]{Staig:2010pn}
\bibinfo{author}{\bibfnamefont{P.}~\bibnamefont{Staig}} \bibnamefont{and}
  \bibinfo{author}{\bibfnamefont{E.}~\bibnamefont{Shuryak}},
  \bibinfo{journal}{Phys. Rev.} \textbf{\bibinfo{volume}{C84}},
  \bibinfo{pages}{034908} (\bibinfo{year}{2011}).

\bibitem[{\citenamefont{Qin et~al.}(2010)\citenamefont{Qin, Petersen, Bass, and
  Muller}}]{Qin:2010pf}
\bibinfo{author}{\bibfnamefont{G.-Y.} \bibnamefont{Qin}},
  \bibinfo{author}{\bibfnamefont{H.}~\bibnamefont{Petersen}},
  \bibinfo{author}{\bibfnamefont{S.~A.} \bibnamefont{Bass}}, \bibnamefont{and}
  \bibinfo{author}{\bibfnamefont{B.}~\bibnamefont{Muller}},
  \bibinfo{journal}{Phys.Rev.} \textbf{\bibinfo{volume}{C82}},
  \bibinfo{pages}{064903} (\bibinfo{year}{2010}), \eprint{1009.1847}.

\bibitem[{\citenamefont{Ma and Wang}(2011)}]{Ma:2010dv}
\bibinfo{author}{\bibfnamefont{G.-L.} \bibnamefont{Ma}} \bibnamefont{and}
  \bibinfo{author}{\bibfnamefont{X.-N.} \bibnamefont{Wang}},
  \bibinfo{journal}{Phys.Rev.Lett.} \textbf{\bibinfo{volume}{106}},
  \bibinfo{pages}{162301} (\bibinfo{year}{2011}), \eprint{1011.5249}.

\bibitem[{\citenamefont{Xu and Ko}(2011{\natexlab{a}})}]{Xu:2010du}
\bibinfo{author}{\bibfnamefont{J.}~\bibnamefont{Xu}} \bibnamefont{and}
  \bibinfo{author}{\bibfnamefont{C.~M.} \bibnamefont{Ko}},
  \bibinfo{journal}{Phys.Rev.} \textbf{\bibinfo{volume}{C83}},
  \bibinfo{pages}{021903} (\bibinfo{year}{2011}{\natexlab{a}}),
  \eprint{1011.3750}.

\bibitem[{\citenamefont{Teaney and Yan}(2011)}]{Teaney:2010vd}
\bibinfo{author}{\bibfnamefont{D.}~\bibnamefont{Teaney}} \bibnamefont{and}
  \bibinfo{author}{\bibfnamefont{L.}~\bibnamefont{Yan}},
  \bibinfo{journal}{Phys.Rev.} \textbf{\bibinfo{volume}{C83}},
  \bibinfo{pages}{064904} (\bibinfo{year}{2011}), \eprint{1010.1876}.

\bibitem[{\citenamefont{Qiu and Heinz}(2011)}]{Qiu:2011iv}
\bibinfo{author}{\bibfnamefont{Z.}~\bibnamefont{Qiu}} \bibnamefont{and}
  \bibinfo{author}{\bibfnamefont{U.~W.} \bibnamefont{Heinz}},
  \bibinfo{journal}{Phys.Rev.} \textbf{\bibinfo{volume}{C84}},
  \bibinfo{pages}{024911} (\bibinfo{year}{2011}), \eprint{1104.0650}.

\bibitem[{\citenamefont{Bhalerao
  et~al.}(2011{\natexlab{a}})\citenamefont{Bhalerao, Luzum, and
  Ollitrault}}]{Bhalerao:2011yg}
\bibinfo{author}{\bibfnamefont{R.~S.} \bibnamefont{Bhalerao}},
  \bibinfo{author}{\bibfnamefont{M.}~\bibnamefont{Luzum}}, \bibnamefont{and}
  \bibinfo{author}{\bibfnamefont{J.-Y.} \bibnamefont{Ollitrault}}
  (\bibinfo{year}{2011}{\natexlab{a}}), \eprint{1104.4740}.

\bibitem[{\citenamefont{Floerchinger and
  Wiedemann}(2011)}]{Floerchinger:2011qf}
\bibinfo{author}{\bibfnamefont{S.}~\bibnamefont{Floerchinger}}
  \bibnamefont{and} \bibinfo{author}{\bibfnamefont{U.~A.}
  \bibnamefont{Wiedemann}} (\bibinfo{year}{2011}), \eprint{1108.5535}.

\bibitem[{\citenamefont{Pang et~al.}(2013)\citenamefont{Pang, Wang, and
  Wang}}]{Pang:2012uw}
\bibinfo{author}{\bibfnamefont{L.}~\bibnamefont{Pang}},
  \bibinfo{author}{\bibfnamefont{Q.}~\bibnamefont{Wang}}, \bibnamefont{and}
  \bibinfo{author}{\bibfnamefont{X.-N.} \bibnamefont{Wang}},
  \bibinfo{journal}{Nucl.Phys.} \textbf{\bibinfo{volume}{A904-905}},
  \bibinfo{pages}{811c} (\bibinfo{year}{2013}), \eprint{1211.1570}.

\bibitem[{\citenamefont{Gale et~al.}(2013{\natexlab{b}})\citenamefont{Gale,
  Jeon, Schenke, Tribedy, and Venugopalan}}]{Gale:2012rq}
\bibinfo{author}{\bibfnamefont{C.}~\bibnamefont{Gale}},
  \bibinfo{author}{\bibfnamefont{S.}~\bibnamefont{Jeon}},
  \bibinfo{author}{\bibfnamefont{B.}~\bibnamefont{Schenke}},
  \bibinfo{author}{\bibfnamefont{P.}~\bibnamefont{Tribedy}}, \bibnamefont{and}
  \bibinfo{author}{\bibfnamefont{R.}~\bibnamefont{Venugopalan}},
  \bibinfo{journal}{Phys.Rev.Lett.} \textbf{\bibinfo{volume}{110}},
  \bibinfo{pages}{012302} (\bibinfo{year}{2013}{\natexlab{b}}),
  \eprint{1209.6330}.

\bibitem[{\citenamefont{Roy et~al.}(2013)\citenamefont{Roy, Mohanty, and
  Chaudhuri}}]{Roy:2012pn}
\bibinfo{author}{\bibfnamefont{V.}~\bibnamefont{Roy}},
  \bibinfo{author}{\bibfnamefont{B.}~\bibnamefont{Mohanty}}, \bibnamefont{and}
  \bibinfo{author}{\bibfnamefont{A.}~\bibnamefont{Chaudhuri}},
  \bibinfo{journal}{J.Phys.} \textbf{\bibinfo{volume}{G40}},
  \bibinfo{pages}{065103} (\bibinfo{year}{2013}), \eprint{1210.1700}.

\bibitem[{\citenamefont{Pang et~al.}(2014)\citenamefont{Pang, Wang, and
  Wang}}]{Pang:2013pma}
\bibinfo{author}{\bibfnamefont{L.}~\bibnamefont{Pang}},
  \bibinfo{author}{\bibfnamefont{Q.}~\bibnamefont{Wang}}, \bibnamefont{and}
  \bibinfo{author}{\bibfnamefont{X.-N.} \bibnamefont{Wang}},
  \bibinfo{journal}{Phys.Rev.} \textbf{\bibinfo{volume}{C89}},
  \bibinfo{pages}{064910} (\bibinfo{year}{2014}), \eprint{1309.6735}.

\bibitem[{\citenamefont{Rybczynski et~al.}(2014)\citenamefont{Rybczynski,
  Stefanek, Broniowski, and Bozek}}]{Rybczynski:2013yba}
\bibinfo{author}{\bibfnamefont{M.}~\bibnamefont{Rybczynski}},
  \bibinfo{author}{\bibfnamefont{G.}~\bibnamefont{Stefanek}},
  \bibinfo{author}{\bibfnamefont{W.}~\bibnamefont{Broniowski}},
  \bibnamefont{and} \bibinfo{author}{\bibfnamefont{P.}~\bibnamefont{Bozek}},
  \bibinfo{journal}{Comput.Phys.Commun.} \textbf{\bibinfo{volume}{185}},
  \bibinfo{pages}{1759} (\bibinfo{year}{2014}), \eprint{1310.5475}.

\bibitem[{\citenamefont{Adare et~al.}(2011)}]{Adare:2011tg}
\bibinfo{author}{\bibfnamefont{A.}~\bibnamefont{Adare}} \bibnamefont{et~al.}
  (\bibinfo{collaboration}{PHENIX Collaboration}) (\bibinfo{year}{2011}),
  \eprint{1105.3928}.

\bibitem[{\citenamefont{Adamczyk et~al.}(2013)}]{Adamczyk:2013waa}
\bibinfo{author}{\bibfnamefont{L.}~\bibnamefont{Adamczyk}} \bibnamefont{et~al.}
  (\bibinfo{collaboration}{STAR Collaboration}), \bibinfo{journal}{Phys.Rev.}
  \textbf{\bibinfo{volume}{C88}}, \bibinfo{pages}{014904}
  (\bibinfo{year}{2013}), \eprint{1301.2187}.

\bibitem[{\citenamefont{Aad et~al.}(2012{\natexlab{b}})}]{ATLAS:2012at}
\bibinfo{author}{\bibfnamefont{G.}~\bibnamefont{Aad}} \bibnamefont{et~al.}
  (\bibinfo{collaboration}{ATLAS Collaboration}), \bibinfo{journal}{Phys.Rev.}
  \textbf{\bibinfo{volume}{C86}}, \bibinfo{pages}{014907}
  (\bibinfo{year}{2012}{\natexlab{b}}), \eprint{1203.3087}.

\bibitem[{\citenamefont{Abelev et~al.}(2013)}]{Abelev:2012ola}
\bibinfo{author}{\bibfnamefont{B.}~\bibnamefont{Abelev}} \bibnamefont{et~al.}
  (\bibinfo{collaboration}{ALICE Collaboration}), \bibinfo{journal}{Phys.Lett.}
  \textbf{\bibinfo{volume}{B719}}, \bibinfo{pages}{29} (\bibinfo{year}{2013}),
  \eprint{1212.2001}.

\bibitem[{\citenamefont{Aad et~al.}(2013{\natexlab{a}})}]{Aad:2012gla}
\bibinfo{author}{\bibfnamefont{G.}~\bibnamefont{Aad}} \bibnamefont{et~al.}
  (\bibinfo{collaboration}{ATLAS Collaboration}),
  \bibinfo{journal}{Phys.Rev.Lett.} \textbf{\bibinfo{volume}{110}},
  \bibinfo{pages}{182302} (\bibinfo{year}{2013}{\natexlab{a}}),
  \eprint{1212.5198}.

\bibitem[{\citenamefont{Chatrchyan
  et~al.}(2013{\natexlab{b}})}]{Chatrchyan:2013nka}
\bibinfo{author}{\bibfnamefont{S.}~\bibnamefont{Chatrchyan}}
  \bibnamefont{et~al.} (\bibinfo{collaboration}{CMS Collaboration})
  (\bibinfo{year}{2013}{\natexlab{b}}), \eprint{1305.0609}.

\bibitem[{\citenamefont{Bozek et~al.}(2013)\citenamefont{Bozek, Broniowski, and
  Torrieri}}]{Bozek:2013ska}
\bibinfo{author}{\bibfnamefont{P.}~\bibnamefont{Bozek}},
  \bibinfo{author}{\bibfnamefont{W.}~\bibnamefont{Broniowski}},
  \bibnamefont{and} \bibinfo{author}{\bibfnamefont{G.}~\bibnamefont{Torrieri}},
  \bibinfo{journal}{Phys.Rev.Lett.} \textbf{\bibinfo{volume}{111}},
  \bibinfo{pages}{172303} (\bibinfo{year}{2013}), \eprint{1307.5060}.

\bibitem[{\citenamefont{Bzdak et~al.}(2013)\citenamefont{Bzdak, Schenke,
  Tribedy, and Venugopalan}}]{Bzdak:2013zma}
\bibinfo{author}{\bibfnamefont{A.}~\bibnamefont{Bzdak}},
  \bibinfo{author}{\bibfnamefont{B.}~\bibnamefont{Schenke}},
  \bibinfo{author}{\bibfnamefont{P.}~\bibnamefont{Tribedy}}, \bibnamefont{and}
  \bibinfo{author}{\bibfnamefont{R.}~\bibnamefont{Venugopalan}}
  (\bibinfo{year}{2013}), \eprint{1304.3403}.

\bibitem[{\citenamefont{Qin and M¨¹ller}(2014)}]{Qin:2013bha}
\bibinfo{author}{\bibfnamefont{G.-Y.} \bibnamefont{Qin}} \bibnamefont{and}
  \bibinfo{author}{\bibfnamefont{B.}~\bibnamefont{M¨¹ller}},
  \bibinfo{journal}{Phys.Rev.} \textbf{\bibinfo{volume}{C89}},
  \bibinfo{pages}{044902} (\bibinfo{year}{2014}), \eprint{1306.3439}.

\bibitem[{\citenamefont{Schenke and Venugopalan}(2014)}]{Schenke:2014zha}
\bibinfo{author}{\bibfnamefont{B.}~\bibnamefont{Schenke}} \bibnamefont{and}
  \bibinfo{author}{\bibfnamefont{R.}~\bibnamefont{Venugopalan}},
  \bibinfo{journal}{Phys.Rev.Lett.} \textbf{\bibinfo{volume}{113}},
  \bibinfo{pages}{102301} (\bibinfo{year}{2014}), \eprint{1405.3605}.

\bibitem[{\citenamefont{Bzdak and Ma}(2014)}]{Bzdak:2014dia}
\bibinfo{author}{\bibfnamefont{A.}~\bibnamefont{Bzdak}} \bibnamefont{and}
  \bibinfo{author}{\bibfnamefont{G.-L.} \bibnamefont{Ma}}
  (\bibinfo{year}{2014}), \eprint{1406.2804}.

\bibitem[{\citenamefont{Aad et~al.}(2013{\natexlab{b}})}]{Aad:2013xma}
\bibinfo{author}{\bibfnamefont{G.}~\bibnamefont{Aad}} \bibnamefont{et~al.}
  (\bibinfo{collaboration}{ATLAS Collaboration}), \bibinfo{journal}{JHEP}
  \textbf{\bibinfo{volume}{1311}}, \bibinfo{pages}{183}
  (\bibinfo{year}{2013}{\natexlab{b}}), \eprint{1305.2942}.

\bibitem[{\citenamefont{Timmins}(2013)}]{Timmins:2013hq}
\bibinfo{author}{\bibfnamefont{A.~R.} \bibnamefont{Timmins}}
  (\bibinfo{collaboration}{ALICE}), \bibinfo{journal}{J.Phys.Conf.Ser.}
  \textbf{\bibinfo{volume}{446}}, \bibinfo{pages}{012031}
  (\bibinfo{year}{2013}), \eprint{1301.6084}.

\bibitem[{\citenamefont{Qin and Muller}(2012)}]{Qin:2011uw}
\bibinfo{author}{\bibfnamefont{G.-Y.} \bibnamefont{Qin}} \bibnamefont{and}
  \bibinfo{author}{\bibfnamefont{B.}~\bibnamefont{Muller}},
  \bibinfo{journal}{Phys.Rev.} \textbf{\bibinfo{volume}{C85}},
  \bibinfo{pages}{061901} (\bibinfo{year}{2012}), \eprint{1109.5961}.

\bibitem[{\citenamefont{Bhalerao
  et~al.}(2011{\natexlab{b}})\citenamefont{Bhalerao, Luzum, and
  Ollitrault}}]{Bhalerao:2011bp}
\bibinfo{author}{\bibfnamefont{R.~S.} \bibnamefont{Bhalerao}},
  \bibinfo{author}{\bibfnamefont{M.}~\bibnamefont{Luzum}}, \bibnamefont{and}
  \bibinfo{author}{\bibfnamefont{J.-Y.} \bibnamefont{Ollitrault}},
  \bibinfo{journal}{Phys.Rev.} \textbf{\bibinfo{volume}{C84}},
  \bibinfo{pages}{054901} (\bibinfo{year}{2011}{\natexlab{b}}),
  \eprint{1107.5485}.

\bibitem[{\citenamefont{Jia}(2013)}]{Jia:2012sa}
\bibinfo{author}{\bibfnamefont{J.}~\bibnamefont{Jia}}
  (\bibinfo{collaboration}{ATLAS Collaboration}), \bibinfo{journal}{Nucl.Phys.}
  \textbf{\bibinfo{volume}{A910-911}}, \bibinfo{pages}{276}
  (\bibinfo{year}{2013}), \eprint{1208.1427}.

\bibitem[{\citenamefont{Qiu and Heinz}(2012)}]{Qiu:2012uy}
\bibinfo{author}{\bibfnamefont{Z.}~\bibnamefont{Qiu}} \bibnamefont{and}
  \bibinfo{author}{\bibfnamefont{U.}~\bibnamefont{Heinz}},
  \bibinfo{journal}{Phys.Lett.} \textbf{\bibinfo{volume}{B717}},
  \bibinfo{pages}{261} (\bibinfo{year}{2012}), \eprint{1208.1200}.

\bibitem[{\citenamefont{Bhalerao et~al.}(2013)\citenamefont{Bhalerao,
  Ollitrault, and Pal}}]{Bhalerao:2013ina}
\bibinfo{author}{\bibfnamefont{R.~S.} \bibnamefont{Bhalerao}},
  \bibinfo{author}{\bibfnamefont{J.-Y.} \bibnamefont{Ollitrault}},
  \bibnamefont{and} \bibinfo{author}{\bibfnamefont{S.}~\bibnamefont{Pal}},
  \bibinfo{journal}{Phys.Rev.} \textbf{\bibinfo{volume}{C88}},
  \bibinfo{pages}{024909} (\bibinfo{year}{2013}), \eprint{1307.0980}.

\bibitem[{\citenamefont{Teaney and Yan}(2014)}]{Teaney:2013dta}
\bibinfo{author}{\bibfnamefont{D.}~\bibnamefont{Teaney}} \bibnamefont{and}
  \bibinfo{author}{\bibfnamefont{L.}~\bibnamefont{Yan}},
  \bibinfo{journal}{Phys.Rev.} \textbf{\bibinfo{volume}{C90}},
  \bibinfo{pages}{024902} (\bibinfo{year}{2014}), \eprint{1312.3689}.

\bibitem[{\citenamefont{Poskanzer and Voloshin}(1998)}]{Poskanzer:1998yz}
\bibinfo{author}{\bibfnamefont{A.~M.} \bibnamefont{Poskanzer}}
  \bibnamefont{and} \bibinfo{author}{\bibfnamefont{S.}~\bibnamefont{Voloshin}},
  \bibinfo{journal}{Phys.Rev.} \textbf{\bibinfo{volume}{C58}},
  \bibinfo{pages}{1671} (\bibinfo{year}{1998}), \eprint{nucl-ex/9805001}.

\bibitem[{\citenamefont{Ollitrault}(1993)}]{Ollitrault:1993ba}
\bibinfo{author}{\bibfnamefont{J.-Y.} \bibnamefont{Ollitrault}},
  \bibinfo{journal}{Phys.Rev.} \textbf{\bibinfo{volume}{D48}},
  \bibinfo{pages}{1132} (\bibinfo{year}{1993}), \eprint{hep-ph/9303247}.

\bibitem[{\citenamefont{Pang et~al.}(2012)\citenamefont{Pang, Wang, and
  Wang}}]{Pang:2012he}
\bibinfo{author}{\bibfnamefont{L.}~\bibnamefont{Pang}},
  \bibinfo{author}{\bibfnamefont{Q.}~\bibnamefont{Wang}}, \bibnamefont{and}
  \bibinfo{author}{\bibfnamefont{X.-N.} \bibnamefont{Wang}},
  \bibinfo{journal}{Phys.Rev.} \textbf{\bibinfo{volume}{C86}},
  \bibinfo{pages}{024911} (\bibinfo{year}{2012}), \eprint{1205.5019}.

\bibitem[{\citenamefont{Petersen et~al.}(2011)\citenamefont{Petersen,
  Bhattacharya, Bass, and Greiner}}]{Petersen:2011fp}
\bibinfo{author}{\bibfnamefont{H.}~\bibnamefont{Petersen}},
  \bibinfo{author}{\bibfnamefont{V.}~\bibnamefont{Bhattacharya}},
  \bibinfo{author}{\bibfnamefont{S.~A.} \bibnamefont{Bass}}, \bibnamefont{and}
  \bibinfo{author}{\bibfnamefont{C.}~\bibnamefont{Greiner}},
  \bibinfo{journal}{Phys.Rev.} \textbf{\bibinfo{volume}{C84}},
  \bibinfo{pages}{054908} (\bibinfo{year}{2011}), \eprint{1105.0340}.

\bibitem[{\citenamefont{Xiao et~al.}(2013)\citenamefont{Xiao, Liu, and
  Wang}}]{Xiao:2012uw}
\bibinfo{author}{\bibfnamefont{K.}~\bibnamefont{Xiao}},
  \bibinfo{author}{\bibfnamefont{F.}~\bibnamefont{Liu}}, \bibnamefont{and}
  \bibinfo{author}{\bibfnamefont{F.}~\bibnamefont{Wang}},
  \bibinfo{journal}{Phys.Rev.} \textbf{\bibinfo{volume}{C87}},
  \bibinfo{pages}{011901} (\bibinfo{year}{2013}), \eprint{1208.1195}.

\bibitem[{\citenamefont{Jia and Huo}(2014{\natexlab{a}})}]{Jia:2014ysa}
\bibinfo{author}{\bibfnamefont{J.}~\bibnamefont{Jia}} \bibnamefont{and}
  \bibinfo{author}{\bibfnamefont{P.}~\bibnamefont{Huo}},
  \bibinfo{journal}{Phys.Rev.} \textbf{\bibinfo{volume}{C90}},
  \bibinfo{pages}{034915} (\bibinfo{year}{2014}{\natexlab{a}}),
  \eprint{1403.6077}.

\bibitem[{\citenamefont{Jia and Huo}(2014{\natexlab{b}})}]{Jia:2014vja}
\bibinfo{author}{\bibfnamefont{J.}~\bibnamefont{Jia}} \bibnamefont{and}
  \bibinfo{author}{\bibfnamefont{P.}~\bibnamefont{Huo}},
  \bibinfo{journal}{Phys.Rev.} \textbf{\bibinfo{volume}{C90}},
  \bibinfo{pages}{034905} (\bibinfo{year}{2014}{\natexlab{b}}),
  \eprint{1402.6680}.

\bibitem[{\citenamefont{Wang and Gyulassy}(1991)}]{Wang:1991hta}
\bibinfo{author}{\bibfnamefont{X.-N.} \bibnamefont{Wang}} \bibnamefont{and}
  \bibinfo{author}{\bibfnamefont{M.}~\bibnamefont{Gyulassy}},
  \bibinfo{journal}{Phys.Rev.} \textbf{\bibinfo{volume}{D44}},
  \bibinfo{pages}{3501} (\bibinfo{year}{1991}).

\bibitem[{\citenamefont{Gyulassy and Wang}(1994)}]{Gyulassy:1994ew}
\bibinfo{author}{\bibfnamefont{M.}~\bibnamefont{Gyulassy}} \bibnamefont{and}
  \bibinfo{author}{\bibfnamefont{X.-N.} \bibnamefont{Wang}},
  \bibinfo{journal}{Comput.Phys.Commun.} \textbf{\bibinfo{volume}{83}},
  \bibinfo{pages}{307} (\bibinfo{year}{1994}), \eprint{nucl-th/9502021}.

\bibitem[{\citenamefont{Deng and Wang}(2010)}]{Wang:2009qb}
\bibinfo{author}{\bibfnamefont{W.-t.} \bibnamefont{Deng}} \bibnamefont{and}
  \bibinfo{author}{\bibfnamefont{X.-N.} \bibnamefont{Wang}},
  \bibinfo{journal}{Phys.Rev.} \textbf{\bibinfo{volume}{C81}},
  \bibinfo{pages}{024902} (\bibinfo{year}{2010}), \eprint{0910.3403}.

\bibitem[{\citenamefont{Huovinen and Petreczky}(2010)}]{Huovinen:2009yb}
\bibinfo{author}{\bibfnamefont{P.}~\bibnamefont{Huovinen}} \bibnamefont{and}
  \bibinfo{author}{\bibfnamefont{P.}~\bibnamefont{Petreczky}},
  \bibinfo{journal}{Nucl.Phys.} \textbf{\bibinfo{volume}{A837}},
  \bibinfo{pages}{26} (\bibinfo{year}{2010}), \eprint{0912.2541}.

\bibitem[{\citenamefont{Cooper and Frye}(1974)}]{Cooper:1974mv}
\bibinfo{author}{\bibfnamefont{F.}~\bibnamefont{Cooper}} \bibnamefont{and}
  \bibinfo{author}{\bibfnamefont{G.}~\bibnamefont{Frye}},
  \bibinfo{journal}{Phys.Rev.} \textbf{\bibinfo{volume}{D10}},
  \bibinfo{pages}{186} (\bibinfo{year}{1974}).

\bibitem[{\citenamefont{Lin et~al.}(2005)\citenamefont{Lin, Ko, Li, Zhang, and
  Pal}}]{Lin:2004en}
\bibinfo{author}{\bibfnamefont{Z.-W.} \bibnamefont{Lin}},
  \bibinfo{author}{\bibfnamefont{C.~M.} \bibnamefont{Ko}},
  \bibinfo{author}{\bibfnamefont{B.-A.} \bibnamefont{Li}},
  \bibinfo{author}{\bibfnamefont{B.}~\bibnamefont{Zhang}}, \bibnamefont{and}
  \bibinfo{author}{\bibfnamefont{S.}~\bibnamefont{Pal}},
  \bibinfo{journal}{Phys.Rev.} \textbf{\bibinfo{volume}{C72}},
  \bibinfo{pages}{064901} (\bibinfo{year}{2005}), \eprint{nucl-th/0411110}.

\bibitem[{\citenamefont{Miller et~al.}(2007)\citenamefont{Miller, Reygers,
  Sanders, and Steinberg}}]{Miller:2007ri}
\bibinfo{author}{\bibfnamefont{M.~L.} \bibnamefont{Miller}},
  \bibinfo{author}{\bibfnamefont{K.}~\bibnamefont{Reygers}},
  \bibinfo{author}{\bibfnamefont{S.~J.} \bibnamefont{Sanders}},
  \bibnamefont{and}
  \bibinfo{author}{\bibfnamefont{P.}~\bibnamefont{Steinberg}},
  \bibinfo{journal}{Ann. Rev. Nucl. Part. Sci.} \textbf{\bibinfo{volume}{57}},
  \bibinfo{pages}{205} (\bibinfo{year}{2007}), \eprint{nucl-ex/0701025}.

\bibitem[{\citenamefont{Zhang}(1998)}]{Zhang:1997ej}
\bibinfo{author}{\bibfnamefont{B.}~\bibnamefont{Zhang}},
  \bibinfo{journal}{Comput.Phys.Commun.} \textbf{\bibinfo{volume}{109}},
  \bibinfo{pages}{193} (\bibinfo{year}{1998}), \eprint{nucl-th/9709009}.

\bibitem[{\citenamefont{Xu and Ko}(2011{\natexlab{b}})}]{Xu:2011fi}
\bibinfo{author}{\bibfnamefont{J.}~\bibnamefont{Xu}} \bibnamefont{and}
  \bibinfo{author}{\bibfnamefont{C.~M.} \bibnamefont{Ko}},
  \bibinfo{journal}{Phys.Rev.} \textbf{\bibinfo{volume}{C83}},
  \bibinfo{pages}{034904} (\bibinfo{year}{2011}{\natexlab{b}}),
  \eprint{1101.2231}.

\bibitem[{\citenamefont{Li and Ko}(1995)}]{Li:1995pra}
\bibinfo{author}{\bibfnamefont{B.-A.} \bibnamefont{Li}} \bibnamefont{and}
  \bibinfo{author}{\bibfnamefont{C.~M.} \bibnamefont{Ko}},
  \bibinfo{journal}{Phys.Rev.} \textbf{\bibinfo{volume}{C52}},
  \bibinfo{pages}{2037} (\bibinfo{year}{1995}), \eprint{nucl-th/9505016}.

\bibitem[{\citenamefont{Zhang et~al.}(2008)\citenamefont{Zhang, Chen, and
  Ko}}]{Zhang:2008zzk}
\bibinfo{author}{\bibfnamefont{B.}~\bibnamefont{Zhang}},
  \bibinfo{author}{\bibfnamefont{L.-W.} \bibnamefont{Chen}}, \bibnamefont{and}
  \bibinfo{author}{\bibfnamefont{C.~M.} \bibnamefont{Ko}},
  \bibinfo{journal}{J.Phys.} \textbf{\bibinfo{volume}{G35}},
  \bibinfo{pages}{065103} (\bibinfo{year}{2008}).

\bibitem[{\citenamefont{Bozek et~al.}(2011)\citenamefont{Bozek, Broniowski, and
  Moreira}}]{Bozek:2010vz}
\bibinfo{author}{\bibfnamefont{P.}~\bibnamefont{Bozek}},
  \bibinfo{author}{\bibfnamefont{W.}~\bibnamefont{Broniowski}},
  \bibnamefont{and} \bibinfo{author}{\bibfnamefont{J.}~\bibnamefont{Moreira}},
  \bibinfo{journal}{Phys.Rev.} \textbf{\bibinfo{volume}{C83}},
  \bibinfo{pages}{034911} (\bibinfo{year}{2011}), \eprint{1011.3354}.

\bibitem[{\citenamefont{Liao and Koch}(2009)}]{Liao:2009ni}
\bibinfo{author}{\bibfnamefont{J.}~\bibnamefont{Liao}} \bibnamefont{and}
  \bibinfo{author}{\bibfnamefont{V.}~\bibnamefont{Koch}},
  \bibinfo{journal}{Phys.Rev.Lett.} \textbf{\bibinfo{volume}{103}},
  \bibinfo{pages}{042302} (\bibinfo{year}{2009}), \eprint{0902.2377}.

\bibitem[{\citenamefont{Han et~al.}(2011)\citenamefont{Han, Ma, Ma, Cai, Chen
  et~al.}}]{Han:2011ys}
\bibinfo{author}{\bibfnamefont{L.}~\bibnamefont{Han}},
  \bibinfo{author}{\bibfnamefont{G.}~\bibnamefont{Ma}},
  \bibinfo{author}{\bibfnamefont{Y.}~\bibnamefont{Ma}},
  \bibinfo{author}{\bibfnamefont{X.}~\bibnamefont{Cai}},
  \bibinfo{author}{\bibfnamefont{J.}~\bibnamefont{Chen}}, \bibnamefont{et~al.},
  \bibinfo{journal}{Phys.Rev.} \textbf{\bibinfo{volume}{C83}},
  \bibinfo{pages}{047901} (\bibinfo{year}{2011}), \eprint{1103.2009}.

\end{thebibliography}

\end{document}